\documentclass[final,3p,twocolumn,nonatbib]{elsarticle}

\usepackage{amssymb}
\usepackage{gensymb}
\usepackage{amsmath}
\usepackage{amsthm}
\usepackage{textcomp}
\usepackage{hyperref}
\usepackage{afterpage}
\usepackage{comment}
\usepackage{graphicx} 
\usepackage{subfigure}
\usepackage{float} 
\usepackage{url}
\usepackage{color} 
\usepackage{enumitem} 
\usepackage{adjustbox}
\usepackage{caption} 
\usepackage{subcaption} 
\usepackage{titlesec}
\usepackage{stfloats}
\usepackage{setspace}
\usepackage{multirow} 
\usepackage{siunitx} 
\usepackage[version=4]{mhchem} 
\usepackage[capitalise, nameinlink, noabbrev]{cleveref}
\usepackage{afterpage} 
\usepackage[backend=biber,style=phys,
articletitle=true,
biblabel=brackets,
terseinits=true,
doi=false,
url=false
]{biblatex}
\AtEveryBibitem{\clearlist{language}}
   
\usepackage{booktabs}
\hypersetup{
    colorlinks=true,
    linkcolor=cyan,
    filecolor=magenta,      
    urlcolor=blue,
    pdftitle={Machine Learning-guided accelerated discovery of structure-property correlations in lean magnesium alloys for biomedical applications},
    pdfpagemode=FullScreen,
    } 
\usepackage{lineno}

\journal{arXiV}

\begin{document}

\newpage
\begin{frontmatter}

\title{Machine Learning-guided accelerated discovery of structure-property correlations in lean magnesium alloys for biomedical applications}

\author[inst1,inst2]{Sreenivas Raguraman\corref{cor1}}

\affiliation[inst1]{organization={Department of Materials Science and Engineering, Johns Hopkins University},
            addressline={3400 N Charles St}, 
            city={Baltimore},
            postcode={21218}, 
            state={MD},
            country={U.S.A}}

\affiliation[inst2]{organization={Hopkins Extreme Materials Institute, Johns Hopkins University},
            addressline={3400 N Charles St}, 
            city={Baltimore},
            postcode={21218}, 
            state={MD},
            country={U.S.A}}

\ead{sragura1@jhu.edu}
\cortext[cor1]{Corresponding authors}

\author[inst3]{Maitreyee Sharma Priyadarshini}

\affiliation[inst3]{organization={Department of Chemical and Biomolecular Engineering, Johns Hopkins University},
            addressline={3400 N Charles St}, 
            city={Baltimore},
            postcode={21218}, 
            state={MD},
            country={U.S.A}}

\author[inst1,inst4]{Tram Nguyen}

\affiliation[inst4]{organization={Translational Tissue Engineering Center, Johns Hopkins School of Medicine},
            addressline={400 N Broadway}, 
            city={Baltimore},
            postcode={21231}, 
            state={MD},
            country={U.S.A}}

\author[inst5]{Ryan McGovern}
\affiliation[inst5]{organization={Department of Biomedical Engineering, Johns Hopkins University},
            addressline={3400 N Charles St}, 
            city={Baltimore},
            postcode={21218}, 
            state={MD},
            country={U.S.A}}
            
\author[inst1]{Andrew Kim}

\author[inst6]{Adam J. Griebel}

\affiliation[inst6]{organization={Research and Development, Fort Wayne Metals Research Products Corp},
            addressline={9609 Ardmore Ave}, 
            city={Fort Wayne},
            postcode={46809}, 
            state={IN},
            country={U.S.A}}
            
\author[inst2,inst3]{Paulette Clancy}
            
\author[inst1,inst2,inst8]{Timothy P. Weihs\corref{cor1}}

\affiliation[inst8]{organization={Department of Mechanical Engineering, Johns Hopkins University},
            addressline={3400 N Charles St}, 
            city={Baltimore},
            postcode={21218}, 
            state={MD},
            country={U.S.A}}

\ead{weihs@jhu.edu}

\begin{abstract}
Magnesium alloys are emerging as promising alternatives to traditional orthopedic implant materials thanks to their biodegradability, biocompatibility, and impressive mechanical characteristics. However, their rapid \textit{in-vivo} degradation presents challenges, notably in upholding mechanical integrity over time. This study investigates the impact of high-temperature thermal processing on the mechanical and degradation attributes of a lean Mg-Zn-Ca-Mn alloy, ZX10. Utilizing rapid, cost-efficient characterization methods like X-ray diffraction and optical, we swiftly examine microstructural changes post-thermal treatment. Employing Pearson correlation coefficient analysis, we unveil the relationship between microstructural properties and critical targets (properties): hardness and corrosion resistance. Additionally, leveraging the least absolute shrinkage and selection operator (LASSO), we pinpoint the dominant microstructural factors among closely correlated variables. Our findings underscore the significant role of grain size refinement in strengthening and the predominance of the ternary Ca$_2$Mg$_6$Zn$_3$ phase in corrosion behavior. This suggests that achieving an optimal blend of strength and corrosion resistance is attainable through fine grains and reduced concentration of ternary phases. This thorough investigation furnishes valuable insights into the intricate interplay of processing, structure, and properties in magnesium alloys, thereby advancing the development of superior biodegradable implant materials.
\end{abstract}



\begin{keyword}
magnesium alloys \sep machine learning \sep corrosion \sep mechanical properties \sep rapid characterization 

\end{keyword}

\end{frontmatter}


\section{Introduction}
\label{sec:Introduction}
Biodegradable magnesium alloys have emerged as a cutting-edge focus of modern materials science and biomedical engineering due to their exceptional mechanical properties and innate ability to degrade within the human body. These attributes make them a compelling choice for medical implants. \cite{witte_history_2010,yang_research_2021} However, to fully harness the potential of biodegradable magnesium alloys, we must gain a comprehensive understanding of how their microstructure changes with thermal treatment and how these transformations affect their mechanical and corrosion properties. \cite{song_control_2007,hofstetter_high-strength_2014,hofstetter_processing_2015} In the context of biodegradable magnesium alloys, striking the right balance between corrosion resistance and deformation resistance becomes all the more critical. These alloys must endure mechanical stresses in their intended applications while gradually degrading as new tissue forms. Achieving this balance is a substantial challenge, as pure magnesium is renowned for its excellent biocompatibility but is susceptible to deformation and rapid corrosion. One standard method to enhance the properties of magnesium is to introduce alloying elements that stimulate the formation of secondary phases. These secondary phases substantially improve the material's mechanical strength, but often at the cost of reduced corrosion resistance, a trade-off of particular relevance to biodegradable magnesium alloys.  \cite{prasadh_current_2022,xue_biodegradable_2022,prasadh_metallic_2022} Given the growing interest in developing magnesium alloys for biomedical purposes, exemplified by the recent FDA approval of the RemeOS screw \cite{whooley_fda_2023}, a deeper understanding of the intricate interplay between microstructural features and material properties is of utmost importance.\\

The impact of microstructural features on material properties is often quantified using sophisticated characterization tools such as TEM, which imposes not only substantial financial costs but also consumes considerable time. In response to these challenges, this study harnesses the power of lab-based X-ray diffraction and optical microscopy to rapidly characterize essential microstructural parameters such as dislocation density, crystallographic texture, intermetallic phase fraction, and grain size for a given Mg alloy.\\

One of the significant challenges in identifying the impact of individual microstructural features lies in their interdependence; they are not isolated entities but rather highly correlated. \cite{wu_dynamic_2020,bhattacharyya_texture_2016} This can be illustrated by the example of Zener particle-pinning \cite{humphreys_grain_1996}, in which grain growth is hindered by secondary phases located on grain boundaries. While the restriction of grain growth by the particles enhances hardness and strength, the particles can also accelerate corrosion. Comprehending and untangling the individual effects of these highly correlated features is difficult.\\

\begin{figure*}  
\centering
\includegraphics[width=1\textwidth]{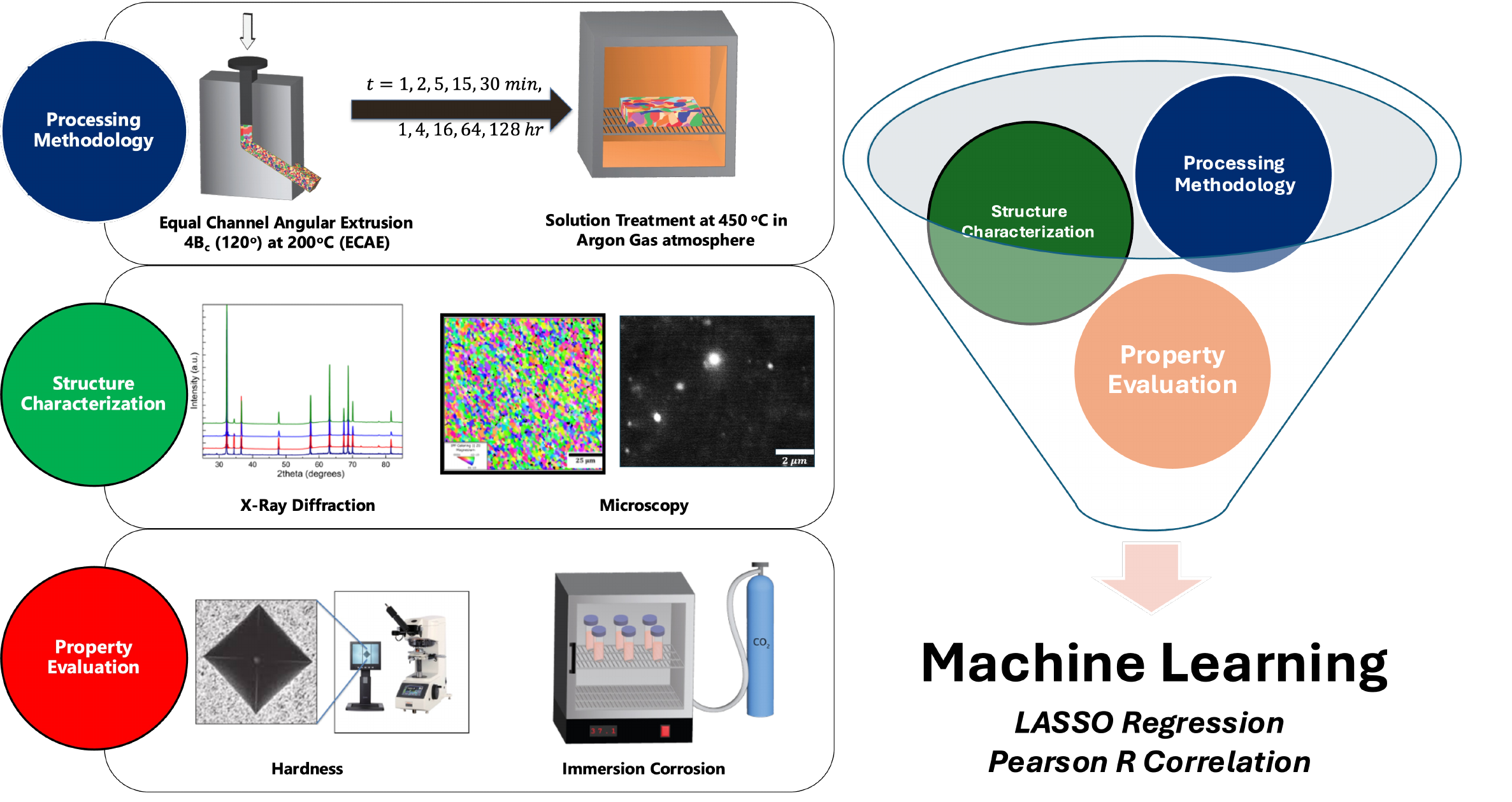}
    \caption{Schematic illustrating the methodologies employed for understanding process-structure-property relationships, including accelerated characterization via XRD and optical microscopy, expedited property assessment through hardness measurements and 1-day immersion studies, and application of machine learning techniques such as Pearson Correlation Coefficient (PCC) analyses~\cite{freedman2007statistics} and LASSO~\cite{tibshirani_regression_1996} for comprehending structure-property correlations.}
    \label{fig: Schematic}
\end{figure*}
Recent advancements in machine learning algorithms have guided a new era of exploration, allowing for an in-depth analysis of individual parameters and the identification of dominant factors that significantly influence properties of interest, as evidenced by numerous studies. \cite{moses_prediction_2023,suh_machine_2023,valipoorsalimi_mechanical_2023} However, most of these investigations have either relied on computationally generated data obtained through calculations or have drawn from experimental data sets scattered across various sources in the literature, lacking a direct one-to-one comparison. Furthermore, many of these studies have concentrated on the chemistry of alloys, with limited attention given to the impact of processing and microstructural features. Recognizing this gap in the literature, our study aims to identify the complex relationships between microstructure, mechanical properties, and corrosion resistance as the microstructure evolves in a dilute biodegradable magnesium alloy, ZX10, as it is heat-treated over a broad range of times at 450 °C. Our study documents the evolution of microstructure, hardness, and \textit{in-vitro} corrosion rates and identifies the dominant microstructural features during different stages of the heat treatment through machine learning, as displayed schematically in \cref{fig: Schematic}.\\

To achieve this, we initiated a simple investigation on a dilute biodegradable magnesium alloy, ZX10, heat-treated over a broad range of times at 450~\degree C. Our study documents the evolution of microstructure, hardness, and \textit{in-vitro} corrosion rates and identifies the dominant microstructural features during different stages of the heat treatment through machine learning, as displayed schematically in \cref{fig: Schematic}.
\section{Materials and methods}
\label{sec: Materials and methods}

\subsection{Thermomechanical Processing}
The ZX10 quaternary alloy, consisting of high-purity Mg and small concentrations of Zn (1 wt\%), Ca (0.3 wt\%), and Mn (0.15 wt\%), was synthesized through a multi-step process. Initially, ingots were created by melting the constituent elements, followed by a homogenization. The ingots were then conventionally extruded at 350°C, using an extrusion ratio of 25:1, resulting in cylindrical rods with a diameter of $\approx$ 12 mm. The extruded rods underwent further processing using the continuous Equal Channel Angular Pressing (cECAP) method, where the extruded rods were subjected to four passes in the $B_c$ route at 300°C, followed by an additional four passes in the $B_c$ route at 200°C through a square die of side 11 mm at an angle of $120\degree$ generating an equivalent strain of 0.67 per pass, as described in Davis \textit{et al.}. \cite{davis_isothermal_2020} Samples processed by cECAP are the starting material for this study and are hereafter referred to as `ECAP'. \\

\begin{figure*}[h!]
    \centering
    \subfigure[]{\includegraphics[width=0.4\linewidth,origin=c]{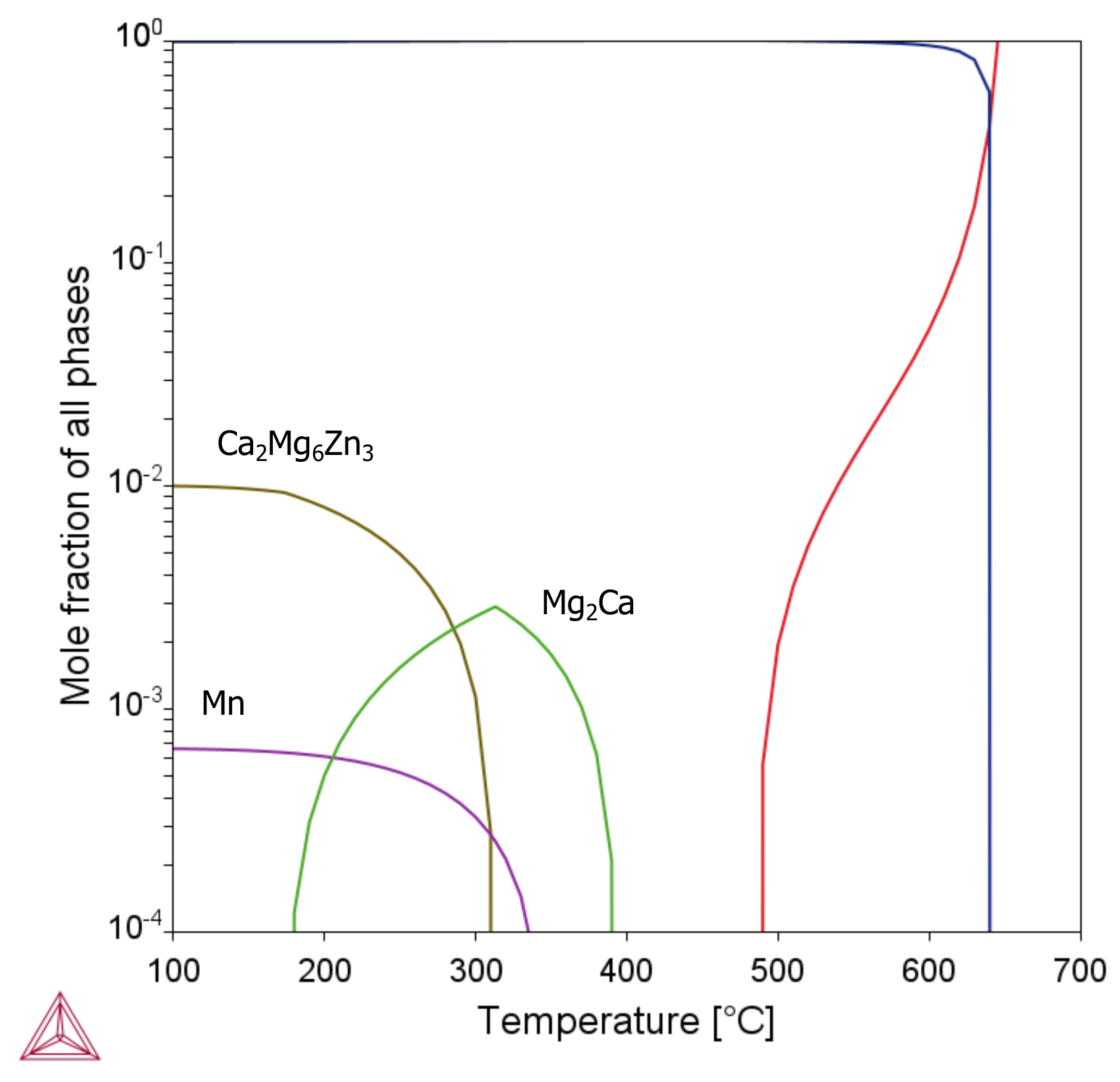} \label{subfig:phase_diagram}}
    \subfigure[]{\includegraphics[width=0.55\linewidth,origin=c]{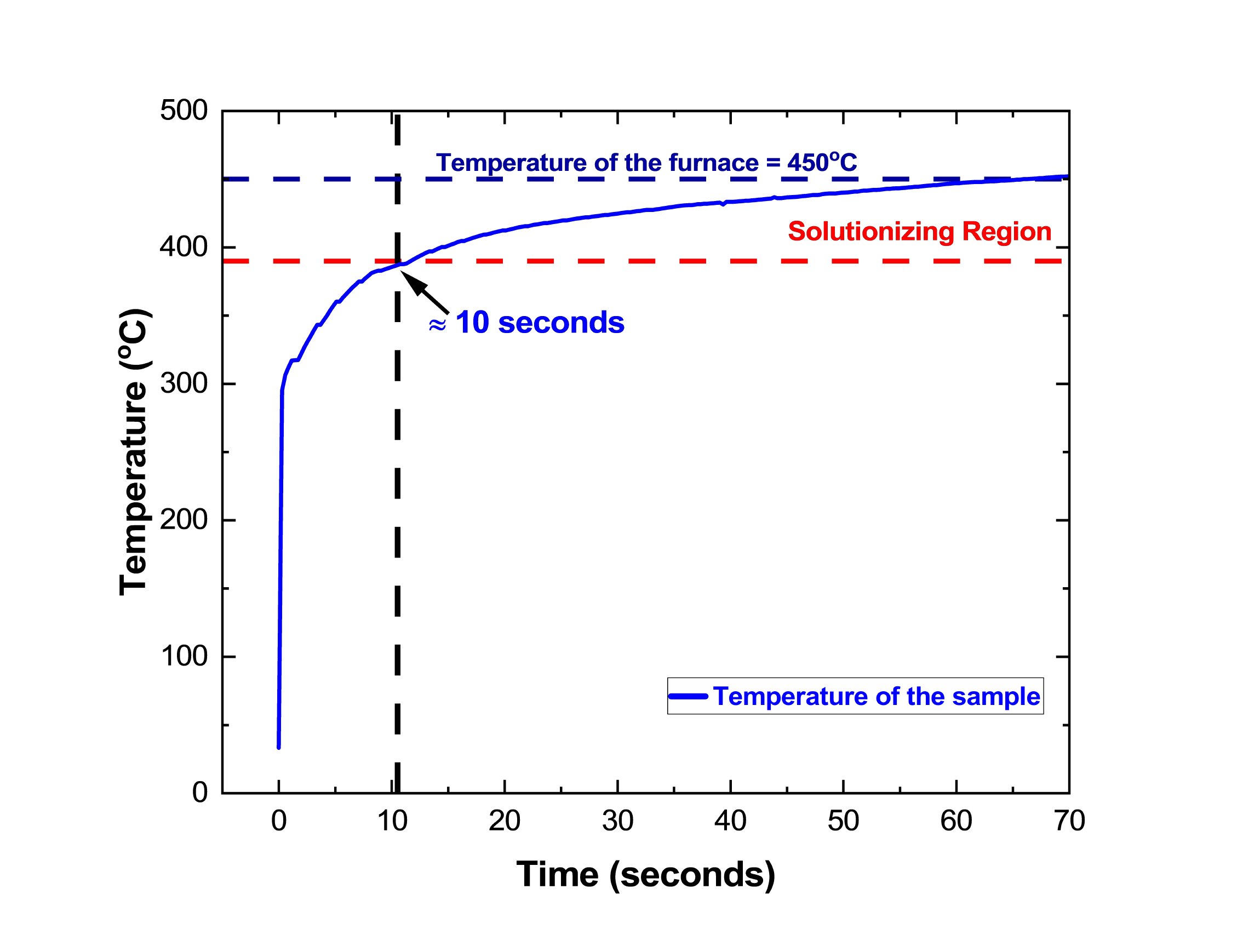} \label{subfig:thermocouple_measurements}}
    \caption{ a) ZX10 Phase Diagram Predicted by ThermoCalc, b) Thermocouple readings of a sample at the start of the solution heat treatment.}
    \label{fig:Processing}
\end{figure*}

Prior to solution heat treatment, samples were precision-cut into 11 by 11 by 1 mm$^3$ sheets using wire electric discharge machining, and the heat-affected regions were removed by polishing with SiC P4000 sandpaper (sourced from Allied HighTech) and a final polishing step with 0.05 $\mu$ m Colloidal Silica Suspension to achieve a mirror-like finish. In addition, ThermoCalc, a CALPHAD-based program, was utilized to calculate the phase diagram of this alloy (as seen in \cref{subfig:phase_diagram}). Based on the predicted phase diagram, the ECAP samples were subjected to solution heat treatment in an Argon gas environment within a Carbolite Gero HTCR5/95 furnace at 450\degree C, where only the $\alpha$-Mg phase is stable. Heat treatments ranged from one minute to a maximum of 128 hours, followed by rapid quenching in water. Moving forward, each sample is designated by the duration at 450°C. For instance, a sample that undergoes a 2-minute solution heat treatment is referred to as ``2 min" sample. A temperature versus time profile for the initial heating is plotted in \cref{subfig:thermocouple_measurements} and shows that samples rise into the solutionizing region, where no secondary phases are stable, within 10 seconds.

\subsection{Characterization and Quantification of Microstructural Parameters}\label{subsec:methods_Quantification_XRD} 
To characterize and quantify microstructural features rapidly, we employed optical microscopy and X-ray diffraction. We also leveraged SEM imaging, SEM-EDS mapping, and SEM-EBSD mapping to verify the consistency of the rapid analyses. To examine the ECAPed and annealed microstructures, samples were etched using a 5\% Nital solution (composed of 5\% Nitric Acid and 95\% methanol) following the ASTM E 407 standard. \cite{e04_committee_practice_nodate} Optical microscopy was performed using a Leica DMi8 Inverted Microscope with LASX software for grain size visualization. JEOL IT700HR InTouchScopeTM SEM and Thermo Scientific Helios 5 UC Focused Ion Dual Beam were employed for imaging and EDS analysis, with EDS data analysis conducted using APEX EDS software. Grain sizes were measured using optical micrographs, scanning electron micrographs, and a MATLAB program, following the intercept method specified in ASTM E112-13. \cite{e04_committee_test_nodate}\\

The mirror-finish samples were subjected to bulk ion beam milling using a Leica EM TIC 3X at 5kV and 3mA for 5-minutes before Electron Backscattered Diffraction (EBSD) studies on a Tescan MIRA 3 GM Field Emission SEM, equipped with Oxford EBSD detector at voltage of 20~kV. Textural and grain size analyses for EBSD were carried out using AZteC Crystal Software by Oxford. We also used AZteC Crystal Software to obtain the Geometrically Necessary Dislocation (GND) density, based on the Weighted Burgers Vector Method developed by Wheeler \textit{et al.} \cite{wheeler_weighted_2009} To mitigate the step size influence on GND measurement and examine relative GND density changes, a constant step size of 0.5 $\mu m$ was employed. The analysis focused on the dislocation density of Burgers vectors with a magnitude of $\frac{1}{2}[11\overline{2}0]$, and the GND density was assessed using a $3\times 3$ pixel kernel size.\\

The SEM-EDS images were initially subjected to thresholding by the Otsu method \cite{otsu_threshold_1979} to distinguish the precipitates from the matrix based on contrast. The representative diameter of each precipitate was then computed by averaging its major and minor axes, which served as the basis for calculating the precipitate size distribution and area fraction.\\

X-ray diffraction (XRD) analysis was conducted per the principle of Bragg's law \cite{cullity_elements_1956} using the Malvern Panalytical Aeris powder X-ray diffractometer, operating at 40~kV and 7.5~mA. 
The instrument was equipped with a Nickel Beta Filter and a Cu X-ray source, and data collection was performed using a step size of 0.0027\degree. 
To ensure data quality, scans were repeated three times and subsequently summed to enhance the signal-to-noise ratio. Data processing, such as background subtraction, removal of K$\alpha_2$ peaks, peak identification, and peak matching, was accomplished using X'pert HighScore software.\\

The Convolutional Multiple Whole Fitting (CMWP) program developed by Ribarik \textit{et al.} was employed to determine the average dislocation density. The standard reference material (SRM 660, LaB$_6$) from the National Institute of Standards and Technology was employed to obtain the instrumental profile function. Due to very minimal alloying content, the broadening of the Mg peaks was assumed to arise mainly from the strain contributions of the dislocations in the Mg phase. The CWMP program has been detailed in \cite{ribarik_mwp-fit_2001,ribarik_correlation_2004,shankar_development_2020,agnew_x-ray_2023}.\\
Peaks exceeding a minimum threshold of 0.01 counts after background subtraction were detected and subjected to PseudoVoigt profile fitting using X'Pert HighScore Software, as detailed in \cite{speakman_profile_2012}. To address the impact of texture on intensity, which can arise from preferential precipitation on specific planes during annealing, multiple peaks with $2\theta$ values spanning the range of 20-60\degree~ were taken into account. \\

We estimated the volume fraction for each constituent phase by calculating the ratio of the integrated area corresponding to an individual phase to the sum of integrated areas for all phases as described in \cite{ehtemam-haghighi_phase_2016} and as illustrated below:
\begin{equation}
      V_{f,a}=\dfrac{A_a}{A_a+A_b+A_c+A_d}
\end{equation}
For texture characterization, we focused on the prismatic  $[10\overline{1}0]$ and $[11\overline{2}0]$ and basal $[0002]$ planes. The degree of texture was quantified using ratios of the integrated peak areas for prismatic planes normalized with respect to the basal plane, as follows:
\begin{equation}
    \dfrac{V_{(10\overline{1}0)}}{V_{(0002)}}=\dfrac{A_{(10\overline{1}0)}}{A_{(0002)}} ,
\hspace{20pt}
\dfrac{V_{(11\overline{2}0)}}{V_{(0002)}}=\dfrac{A_{(11\overline{2}0)}}{A_{(0002)}}
\end{equation}
\subsection{Microhardness evaluation}
Microhardness measurements were conducted utilizing a 200 gf load on a LECO AMH55 Hardness Tester. Vickers hardness values were derived from the indentation size through the 'Cornerstone' software. Each hardness data point represents a total of at least ten measurements. In accordance with DIN-ISO 6507 guidelines \cite{noauthor_din_nodate}, the micro-indents were spaced at distances equal to six times the average indent width. 
\subsection{Bio-corrosion Evaluation} 
The square samples measuring $11 \times 11 \times 1 \, \text{mm}^3$ were polished on all sides with P4000 SiC paper prior to immersion in Earle's Balanced Salt Solution at 5\% $\text{CO}_2$ and a temperature of 37.1°C, simulating in-vivo conditions for 24 hours. Immersion bio-corrosion tests were conducted with a sample surface area to solution volume ratio of $0.2 \, \text{mL/mm}^2$ per ASTM G31-72 standard \cite{j01_committee_guide_nodate}. The testing took place within a Heraeus Heracell \ce{CO2} 150 incubator. Following the immersion tests, the corroded samples were treated with a solution composed of $(200 \, \text{g} \, \text{CrO}_3 + 10 \, \text{g} \, \text{AgNO}_3 + 20 \, \text{g} \, \text{Ba(NO}_3)_2$ dissolved in $1l$ of deionized water, following the guidelines set forth by ASTM G1 \cite{g01_committee_practice_nodate} to remove corrosion products. Mass and pH measurements were acquired before and after the corrosion testing utilizing a weighing scale (Hanchen Electronic Analytical Balance, 0.1~mg, Digital Scale) and benchtop pH meter (Accumet AB150, Thermo Fisher Scientific, MA, USA). The pH level was maintained below 8. The biocorrosion rate of each sample was calculated by the weight loss measured in an immersion period of 24 hours according to the equation in \cite{j01_committee_guide_nodate}:
\begin{equation}
    \text{Corrosion Rate} = \dfrac{K\times W}{A \times T \times \rho} \hspace{5pt} mm/yr
\end{equation}
where $K$ is a constant ($8.76 \times 10^4$),
$W$ is the weight loss in the unit of gram, $A$ is the exposed sample surface area in the unit of cm$^2$, T is the time of exposure in the unit of hours, and $\rho$ is the sample density in the unit of g/cm$^3$. The calculated density of $1.77\pm 00.26$, obtained using a helium gas pycnometer (Micromeritics AccuPyc II 1340) and a microbalance (Mettler Toledo Model XS3DU), were employed to estimate corrosion rates. For this study, we focused on corrosion over a single day, given that corrosion rates are usually at their highest initially and tend to decrease over time. 
\subsection{Machine Learning-driven Analysis} \label{subsec:methods_ML}
We apply two machine learning techniques to understand the correlations between the microstructural features and to identify the most significant microstructural features that affect the corrosion rate and hardness. 
To obtain correlations between the microstructural features, we compute the Pearson R correlation coefficients (PCC)~\cite{freedman2007statistics} using the following equation:
\begin{equation}
r_{xy}={\frac {\sum _{i=1}^{n}(x_{i}-{\bar {x}})(z_{i}-{\bar {z}})}{{\sqrt {\sum _{i=1}^{n}(x_{i}-{\bar {x}})^{2}}}{\sqrt {\sum _{i=1}^{n}(z_{i}-{\bar {z}})^{2}}}}},
\end{equation}
where $x$ and $z$ represent two different microstructural features, $x_i$ and $z_i$ represent data observations for the given microstructural feature, and $\bar {x}$ and $\bar {z}$ represent the mean of the observations for $x$ and $z$. 
The PCC is computed pairwise for the microstructural features and is a statistical measure of the linear correlation between two data sets. 
A positive PCC value indicates that the values $x$ increases as the value of $y$ increases, and a negative PCC indicates that the value of $x$ decreases as the value of $y$ increases. 
We also compute the PCC between each microstructural feature and the targets - hardness and corrosion rate. 
This analysis provides an understanding of how each microstructural feature impacts the strengthening and degradation of the Mg alloy. \\

Following the correlation analysis, we performed feature selection using LASSO. \cite{tibshirani_regression_1996} 
LASSO builds a linear model based on the assumption that the model coefficient vector ($\beta$) is sparse. This implies that only some of the input variables are selected to create the linear model.
The following objective function is solved by LASSO: 
\begin{equation}
\min _{\beta \in \mathbb {R} ^{p}}\left\{{\frac {1}{N}}\left\|y-x\beta \right\|_{2}^{2}+\lambda \|\beta \|_{1}\right\},
\end{equation}
where $N$ is the number of data points, $\beta$ is the coefficient vector, $x$ represents the microstructural features data, $y$ represents the targets, hardness, and corrosion rate, and $\lambda$ represents a tunable hyperparameter that controls the sparsity of our linear model.
In our implementation for this work, we perform a grid search on the $\lambda$ parameter and select the one that gives the highest accuracy model based on leave-one-out-cross-validation (LOOCV). 
We use the mean absolute error and R$^2$ score as the accuracy measure for cross-validation. 

\section{Results and Discussion}
\label{sec:Results and Discussion}
\subsection{Variation in Hardness and Corrosion on Solution Heat Treatment}

\begin{figure}
\centering
\includegraphics[width=0.8\linewidth]{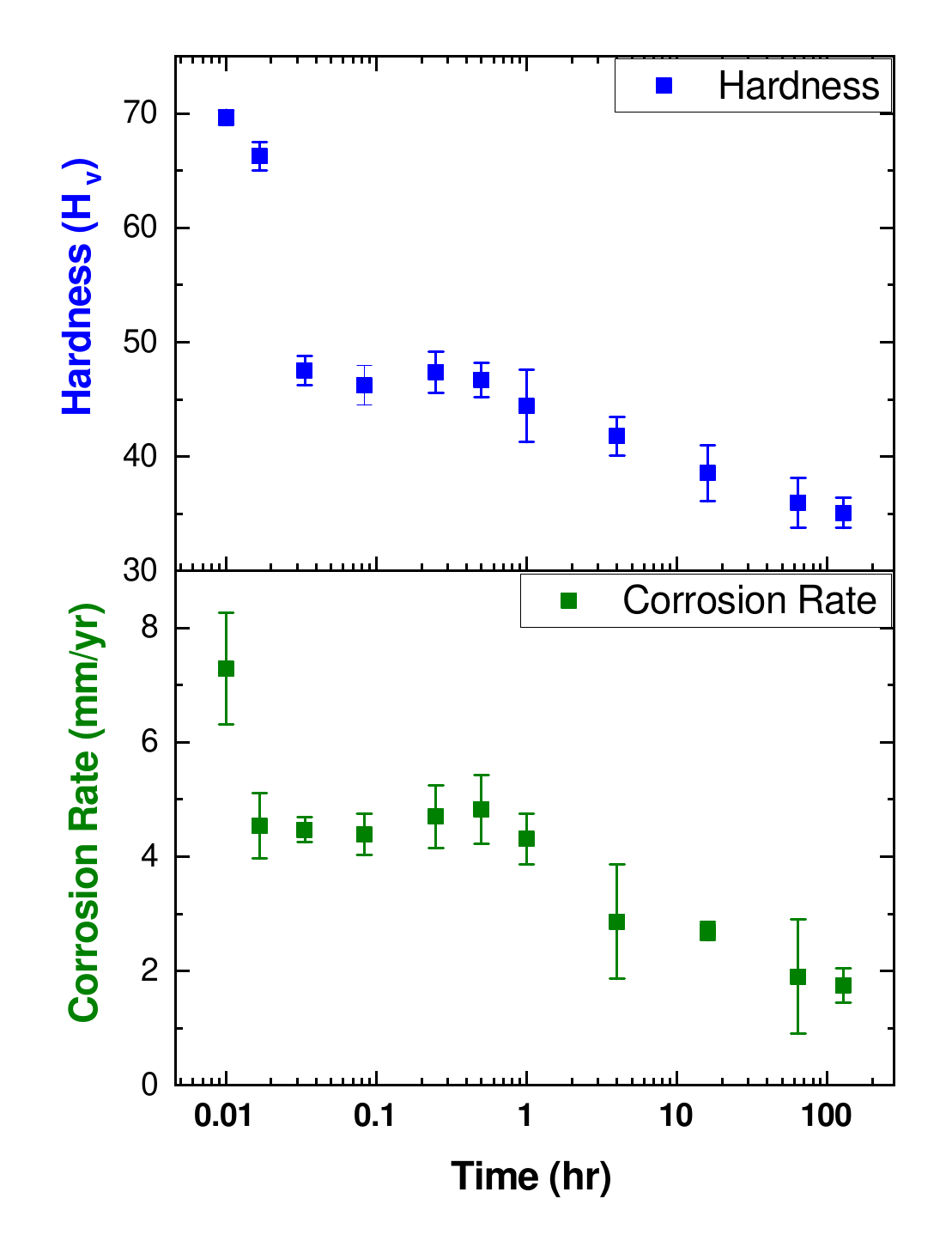}
    \caption{Evolution in hardness and corrosion rate with time reveals the loss of strength and reduced corrosion rate over time.}
    \label{fig: Property Evolution}
\end{figure}

As depicted in \cref{fig: Property Evolution},  the initial minute of heat treatment produces minimal impact on hardness; however, there was a sharp decrease (by 30\%) after 2-minutes. The hardness then stabilized for approximately 30-minutes before gradually declining, resulting in a cumulative reduction of 50\% by 128 hours. In contrast, the corrosion rate experienced a rapid decline (40\%) within the first minute, stabilized for the subsequent 30-minutes, and then gradually decreased until a total reduction of 4X was observed at 128 hours. Notably, such substantial property changes occurring within brief time intervals (1-2 minutes) are unusual. \\

It was intriguing to note the combination of high strength and relatively favorable corrosion rates in the 1 min sample. This observation suggests that high-temperature annealing for shorter durations can enhance the corrosion behavior of heavily deformed samples. These observations also underscore the remarkable changes in material properties without alterations in chemistry, emphasizing the critical role of processing on properties. The underlying mechanisms driving these transformations lie within the microstructure, underlining the necessity for ongoing monitoring of microstructural changes to attain optimal material properties.
\subsection{Tracking material microstructure rapidly following heat treatment}

\begin{figure*}
\centering
\includegraphics[width=0.8\linewidth]{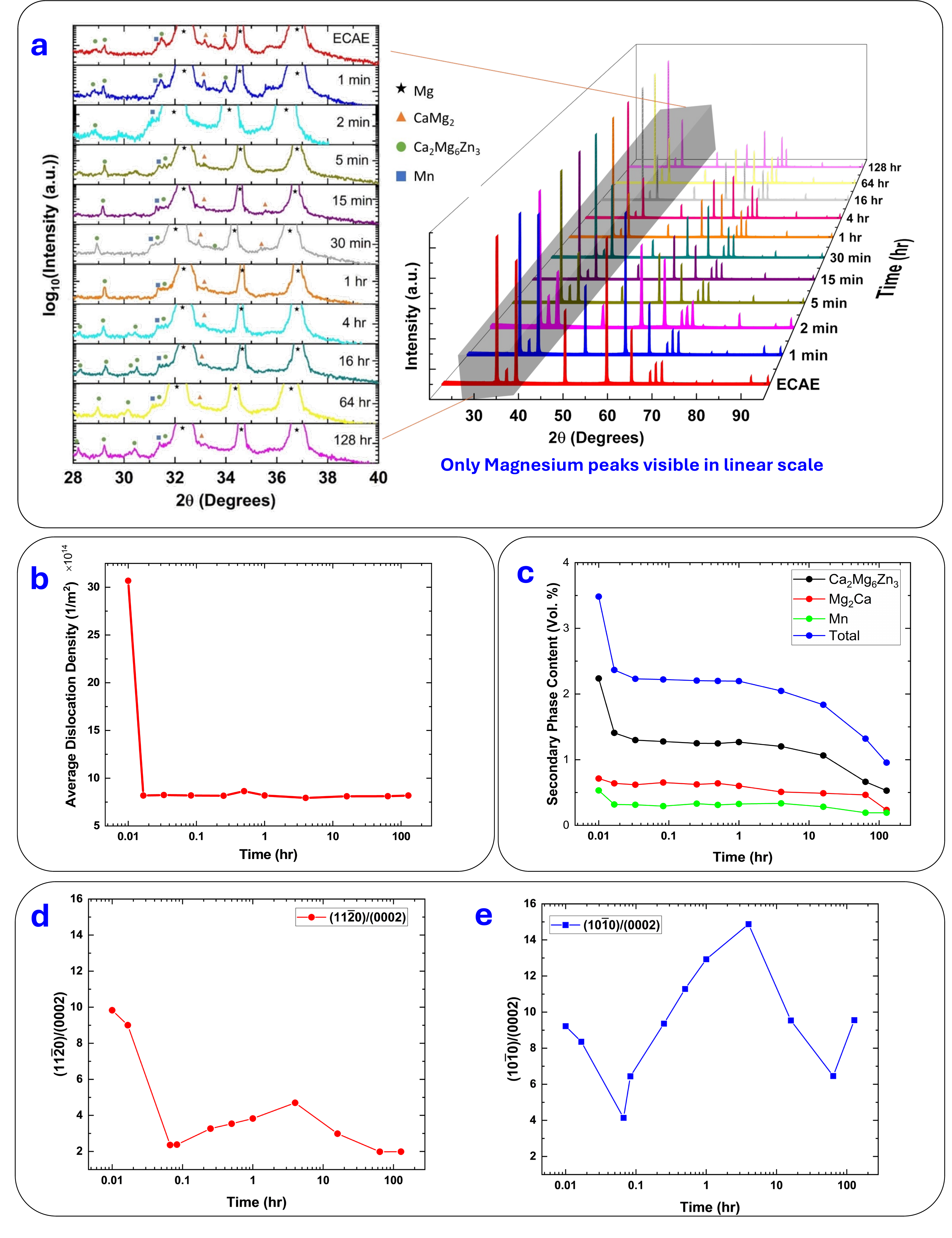}
    \caption{\textbf{Accelerated Microstructural characterization from XRD:} a) XRD patterns revealing the Mg peaks (in linear scale) and presence of precipitates (in logarithm scale) b) Dislocation density obtained from CMWP analyses revealing dislocation annihilation and plateauing post-1-minute annealing; c) Variation in precipitate content obtained from XRD revealing dissolution of precipitates; Texture ratios from XRD showing the variation of (d) (10$\overline{1}$0)/(0002) and (e) (11$\overline{2}$0)/(0002) over time.}
    \label{fig: XRD_data}
\end{figure*}

After subjecting the samples to heat treatments and quenching, we rapidly characterize four critical microstructural features pertinent to hardness and corrosion using X-ray diffraction (XRD) and optical microscopy (OM) methods, which offer advantages in terms of minimal sample preparation, time efficiency, and cost-effectiveness. These features include dislocation density, crystallographic texture, precipitate volume, and grain size. \cite{raguraman_impact_2024,bahmani_corrosion_2022} While the XRD-derived values provide averaged measurements and may not capture local variations in dislocation densities or increases in particle size beyond $200$ $nm$, they serve as valuable tools for explaining the observed fluctuations in mechanical and corrosion properties as depicted in \cref{fig: Property Evolution}. 
XRD patterns and optical microstructures of all the conditions are plotted in \cref{fig: XRD_data,fig: GG}, respectively. 

\subsubsection{Variation in Dislocation Density} \label{subsubsec:dislocation_density}
\begin{figure*}
\centering
\includegraphics[width=0.7\linewidth]{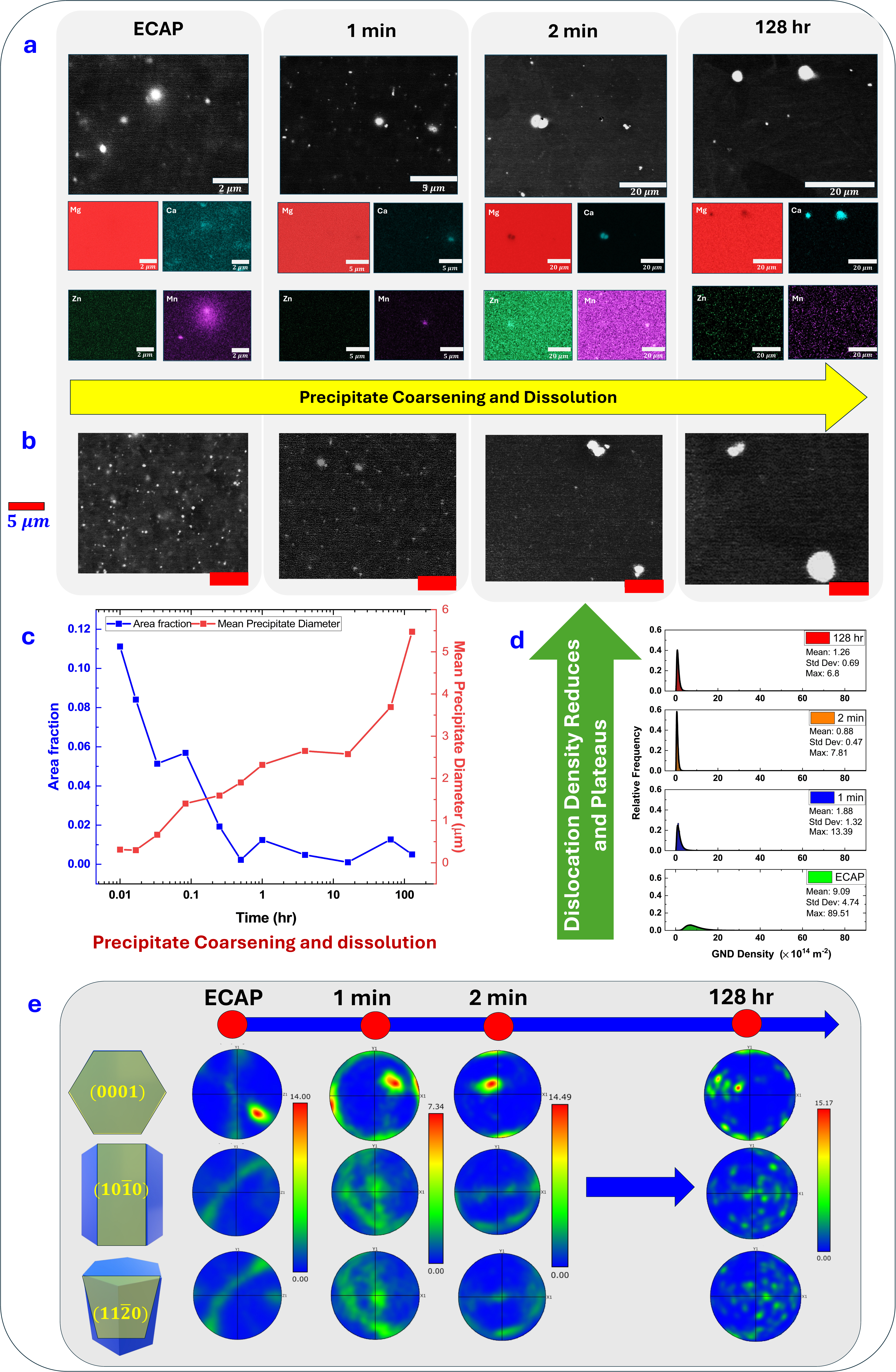}
    \caption{\textbf{Microstructural Evolution validation using SEM}: a) SEM-EDS Maps revealing the presence of different precipitates; b) BSE images showing precipitate dissolution and coarsening over time; c) Variation in area fraction and mean diameters of precipitates obtained from SEM showing precipitate dissolution and coarsening over time; d) Log-normal distributions of GND Density revealing dislocation annihilation and plateauing post-1-minute annealing validating the trend from XRD; g) EBSD Pole figures revealing the texture variation over time}
    \label{fig: Microstructural Evolution}
\end{figure*}

Based on CMWP analyses as demonstrated in \cref{fig: XRD_data}\textcolor{blue}{(b)}, the ECAP sample has a high dislocation density exceeding 3$\times 10^{15}$ m$^{-2}$ but drops tenfold after just a one-minute exposure to $450\degree$C, indicative of significant dislocation annihilation.
This dramatic drop is supported by EBSD measurements, as demonstrated by the geometrically necessary dislocation (GND) density in \cref{fig: Microstructural Evolution}\textcolor{blue}{(d)}. A noteworthy aspect is the consistent log-normal distribution of GND density within the heat-treated samples.\\

Previous studies \cite{kapoor_influence_2020,zemkova_influence_2018,wu_dislocation_2022} have predominantly examined variations in dislocation density over longer time scales, typically spanning 20 minutes or more. Given our study utilizes far shorter time durations, we conducted \textit{in-situ} temperature measurements with thermocouples to ensure samples reach the annealing temperature rapidly, as shown in \cref{subfig:thermocouple_measurements}. Note that the samples reach the solutionizing region within 10 seconds, so even a brief one-minute annealing period provides sufficient exposure to high temperatures for the annihilation of dislocations.

\subsubsection{Crystallographic Texture Evolution}
\label{subsec: texture}
In terms of crystallographic texture, we primarily considered three planes for our study: the basal $(0002)$ plane and the prismatic $(10\overline{1}0)$ and $(11\overline{2} 0)$ planes, which have been identified as significant influencers of corrosion behavior as per literature. It is worth noting that prismatic planes corrode 18-20 times faster than basal planes. \cite{gerashi_effect_2022, bahmani_corrosion_2022} Thus, texture ratios were calculated using integrated areas from XRD scans as detailed in \cref{subsec:methods_Quantification_XRD}, and both integrated area ratios $(10\overline{1}0)/(0002)$ and $(11\overline{2}0)/(0002)$ show similar trends over time as seen in \cref{fig: XRD_data}\textcolor{blue}{(d)} and \textcolor{blue}{(e)}. The XRD intensity of the $(10\overline{1}0)$ plane remains relatively constant from the first minute onwards, whereas the intensities of  the $(11\overline{2}0)$ and $(0002)$ planes fluctuate in response to the heat treatment duration. The most pronounced variations occur during the second minute, which coincides with the onset of grain growth. \\

The EBSD pole figures reveal a similar trend while offering intriguing additional insights. The texture notably weakens from the ECAP to the 1-minute sample. In the ECAP sample, the basal plane texture tilts approximately $45\degree$ to the normal. Conversely, we observe a wider range of texture variation in the one-minute sample, spanning from $25\degree$ to $65\degree$ to the normal. However, a significant texture strengthening occurs in the 2-minute sample, with the basal pole aligning more closely towards the center of the pole figure. As subsequent sections will explain, the dissolution of precipitates leads to an increase in grain size from 1-minute to 2-minutes, consequently altering the texture. By the 128-hour mark, the basal texture, $(0001)$,  exhibits a more dispersed pattern, aligning closer to the $0-45\degree$ range along the x-axis, as illustrated in \cref{fig: Microstructural Evolution}\textcolor{blue}{(e)}. 

\subsubsection{Precipitate Evolution}

In this alloy, three common types of precipitates were observed: Ca$_2$Mg$_6$Zn$_3$, Mg$_2$Ca, and Mn, as evidenced by both XRD (\cref{fig: XRD_data}\textcolor{blue}{(c)}) and SEM-EDS analyses (\cref{fig: Microstructural Evolution}\textcolor{blue}{(a-c)}). This observation aligns with the predicted phase diagram by ThermoCalc, illustrated in \cref{subfig:phase_diagram}. At $450\degree \text{C}$, all three phases are thermodynamically unstable, and the stable phase is a solid solution. However, solutionizing completely requires significant time.\\

While determining phase volume fraction using integrated area measurements from XRD (\cref{fig: XRD_data}), the influence of texture on preferential precipitation was deemed negligible. As expected, the calculated volume fraction of all three phases decreased with annealing time, as illustrated in \cref{fig: Microstructural Evolution}\textcolor{blue}{(b)}.
Within the initial minute of exposure at 450°C, the overall precipitate fraction decreased by approximately 30\%. Subsequently, from 1-2 minutes, there was a decrease of around 6\%, followed by stepwise declines in the precipitate fraction, each below 1\%, until the 4th-hour mark, when more pronounced decreases occur over time, resulting in close to a 70\% reduction in precipitate content.\\

While XRD measurements do not provide detailed insights into precipitate size and distribution, they facilitate the identification of precipitate dissolution and differential dissolution rates among precipitates, which are challenging to obtain through optical microscopy or SEM. On the other hand, SEM-EDS maps provided an understanding of precipitate morphology within its resolution limit and also confirmed the dissolution of precipitates over time, corroborating our XRD analyses. Image analysis of backscattered electron (BSE) images also revealed that precipitate diameter increased over time while area fraction decreased, as seen in \cref{fig: Microstructural Evolution}\textcolor{blue}{(b)} and \cref{fig: Microstructural Evolution}\textcolor{blue}{(c)}.\\

Upon closer examination of individual precipitate variation over time, the initial drop in precipitate content within the first minute is attributed primarily to the ternary Ca$_2$Mg$_6$Zn$_3$ phase, exhibiting close to a 40\% decrease. This substantial reduction can be attributed to the high mobility of Zn atoms in the Mg matrix, as calculated from diffusion coefficient data obtained from \cite{zhong_comprehensive_2020}. Zn atoms display a significantly higher interdiffusion coefficient (9.24 $\mu m^2/s$) at 450\degree C compared to Ca atoms (0.19 ${\mu m}^2/s$) and Mn (0.001 ${\mu m}^2/s$), as well as Mg's self-diffusion coefficient (0.03 $\mu m^2/s$). The data also explains the slower dissolution of the Mg$_2$Ca phase and minimal change in the Mn phase.\\

The dissolution of precipitates within such short time scales, as observed in the first minute, appears to be an unexplored phenomenon in past literature. Most studies \cite{nie_precipitation_2012,zhang_dissolution_2015,mohammadi_zerankeshi_effects_2022,liu_influence_2022,he_quantitative_2017} typically focus on precipitate evolution starting at 30-minutes or longer. However, this study underscores the importance of monitoring microstructural features at smaller time scales, particularly at elevated temperatures, in part because short processing times are advantageous to industry.

\subsubsection{Grain Size Variation}

\begin{figure*}
\centering
\includegraphics[width=0.95\linewidth]{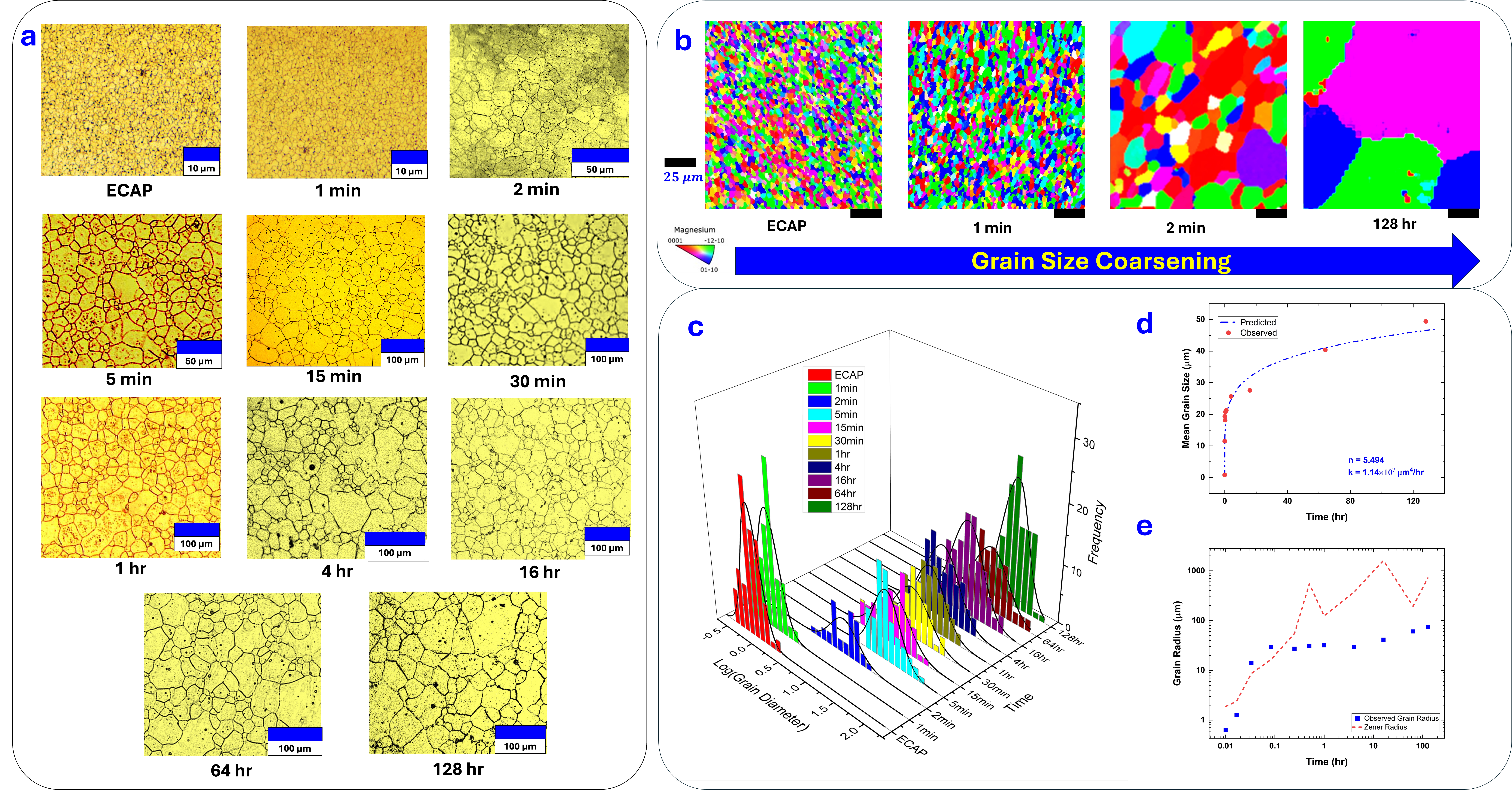}
    \caption{\textbf{Grain Size Analysis:} a) Optical Microstructures revealing grain size coarsening over time; b) Inverse Pole Figures of ECAP, 1 min, 2 min, and 128 hr samples confirming the rapid grain growth; c) Grain size distributions as a function of condition revealing abnormal grain growth; d) The variation in the width of grain diameter over time, and the fit as per \cref{GGeqn} denoting the abnormal grain growth; e) Variation in 3D grain radius as a function of time, with a fit per the Zener model (\cref{Zenereqn}) revealing the abnormal grain growth in the 2$^{nd}$ min and 5$^{th}$ min samples, beyond which the grain growth is normal.}
    \label{fig: GG}
\end{figure*}

Our investigation into grain size primarily relied on optical microscopy (\cref{fig: GG}\textcolor{blue}{(a)}), complemented by EBSD grain size measurements (\cref{fig: GG}\textcolor{blue}{(b)}) at specific time intervals to validate our findings. Initially, the ECAP samples exhibited highly refined grains, averaging approximately 800 $nm$ in size. Despite the observed dislocation annihilation, a one-minute annealing process did not induce significant changes in grain size. However, as the heat treatment extended to 128 hours, the grain size increased significantly by 60-fold, reaching nearly 50 $\mu m$ towards the end of the process.\\

Of particular note is the ten-fold increase in grain size observed from the first minute to the second minute, a phenomenon also evident in the texture variations illustrated in \cref{fig: Microstructural Evolution}\textcolor{blue}{(e)} and \textcolor{blue}{5(i)}. Such substantial changes in microstructure tend to be under-reported in the literature that typically considers longer time periods. \cite{chen_mechanisms_2022,panda_texture_2019}\\

To understand this rise in grain growth behavior, we employed the well-established grain growth model \cite{bhattacharyya_texture_2016}:
\begin{equation}
d^n+d_o^n=kt,   
\label{GGeqn}
\end{equation}
where $k$ is the grain growth exponent, $t$ is the time, $d_o$ is the initial grain size, and $d$ is the grain size at time, $t$. Fitting the model yields a grain growth exponent, $n \approx 5.5$, that is a substantial deviation from the value proposed by Burke and Turnbull (n = 2) \cite{burke_recrystallization_1952} as seen in \cref{fig: GG}\textcolor{blue}{(d)}. This deviation points to the occurrence of abnormal grain growth, a phenomenon characterized by the sudden emergence of unusually large grains within a matrix of uniformly sized grains. An analysis in Supplementary \cref{tab: Grain Growth Variations} reveals the presence of island grains, characterized by grains with double the average grain size, a characteristic feature of abnormal grain size. This phenomenon is not uncommon and has been well-documented in previous studies in magnesium alloys. \cite{chen_mechanisms_2022,bhattacharyya_texture_2015,bhattacharyya_texture_2016,pei_normal_2020} The grain growth exponent observed in this study is consistent with findings from these previous studies, where $n$ typically falls within the range of 2 to 7.\\

This grain growth model assumes boundary curvature drives growth, the absence of a drag force on the boundary due to particles or solutes, and isotropic grain boundary energies and mobilities. Given this alloy contains second-phase particles, we investigated grain size distributions over time, which revealed deviations from self-similarity, a hallmark of abnormal grain growth as described by Humphreys \cite{humphreys_grain_1996} and seen in \cref{fig: GG}\textcolor{blue}{(c)}. A high value of \(n\) points to grain growth stagnation, likely due to second-phase particles exerting a drag force on the boundaries, effectively pinning the microstructure. As illustrated in Supplementary \cref{fig: N-GG}, the normalized grain distribution and the variation in the width of the normalized distribution exhibit specific patterns. The normalized grain size distributions remain log-normal throughout, except in the 2 min sample as revealed in Supplementary \cref{tab:Fit-Log-normal}. Notably, the breadth of the distribution does not change significantly under any of the annealing conditions; instead, it oscillates. This observation suggests that while some grains began growing at a much faster rate than others, broadening the distribution, the finer grains rapidly caught up, preventing distributions from becoming bimodal. \cite{bhattacharyya_texture_2015,bhattacharyya_texture_2016} This transient nature of abnormal grain growth has been observed in previous experiments and simulations. \cite{burke_recrystallization_1952, rollett_simulation_1989,srolovitz_computer_1985,bruno_grain_1995,humphreys_grain_1996}\\

The most commonly cited explanation for abnormal grain growth revolves around the coarsening of particles that effectively pin the grain boundaries, as explained well in \cite{bhattacharyya_texture_2016}. The classical approach to understanding the impact of pinning particles on grain growth is to employ the Zener model \cite{manohar_five_1998}:
\begin{equation}
    R_c = \frac{4r}{3f}, 
\label{Zenereqn}
\end{equation}
where \(R_c\) is the limiting grain radius, \(r\) is the radius of the pinning particles, and \(f\) is the volume fraction of the particles. This equation, while based on several simplifying assumptions like spherical particles and grains and randomly distributed particles with no preferential arrangement at grain boundaries, has been applied here to provide a semi-quantitative comparison between observed and predicted grain sizes.\\

Our observations (\cref{fig: GG}\textcolor{blue}{(e)}) indicate that in the ECAP and 1-minute samples, the observed grain radius remains significantly below the calculated critical radius, hereafter referred to as the Zener radius. There appear to be two distinct regions of normal grain growth separated by a segment of abnormal grain growth (2 min and 5 min): the first region encompasses ECAP and 1-minute, and the second region starts from 15-minutes all the way to 128 hours, aligning with the trend visible in the grain size distribution plot in \cref{fig: GG}\textcolor{blue}{(d)}.\\

This unusual behavior in the 2-minute sample is attributed to the significant dissolution of the ternary phases in the first minute, which leads to an approximately ten-fold increase in average grain size. Since these ternary particles may be heterogeneous in size and distribution, pinning of grain boundaries and grain growth may also be heterogeneous, resulting in a bimodal distribution as revealed in Supplementary \cref{tab:Fit-Log-normal}. However, by 5-minutes, many of these particles dissolve, contributing to a nearly 8-fold rise in grain size from 2-5 minutes and log-normal distributions of grain sizes. Subsequently, the grain growth stabilizes and follows normal grain growth behavior. This study reveals the highly correlated nature of microstructural features, wherein the presence of secondary phases correlates with and appears to control grain growth and texture evolution.

\subsection{Correlations between Microstructural Features}

\begin{figure*}
    \centering
    \includegraphics[width=1\linewidth]{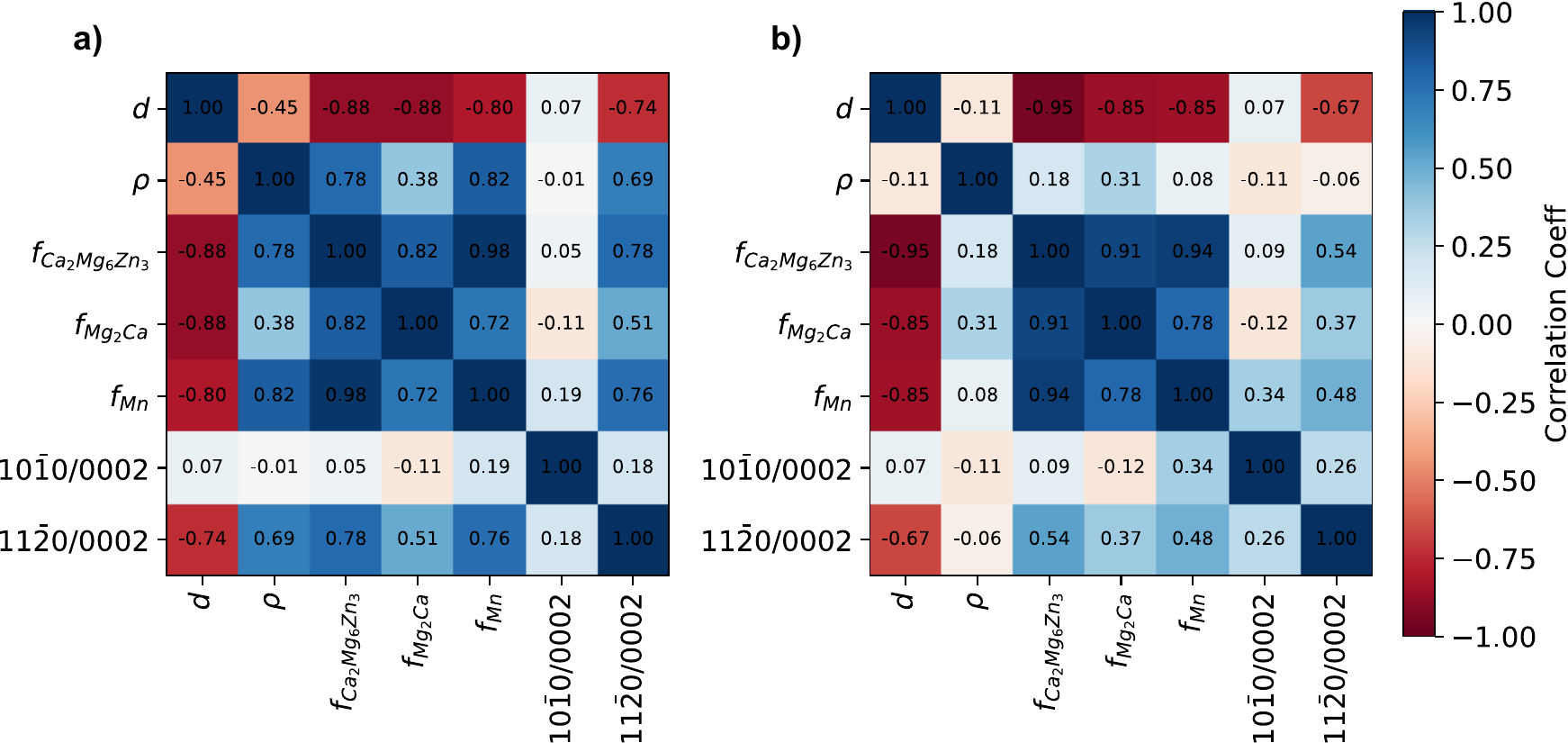}
    \caption{Correlations between the microstructural features (a) with ECAP and (b) without ECAP sample.}
    \label{fig:Correlations}
\end{figure*}
\cref{fig:Correlations} displays the Pearson R correlation coefficients (PCC)~\cite{freedman2007statistics} for the microstructural features characterized in our study. The left matrix in \cref{fig:Correlations}\textcolor{blue}{(a)} presents results obtained while including the ECAP sample before any solution treatment, while the right matrix \cref{fig:Correlations}\textcolor{blue}{(b)} shows correlation coefficients after excluding data for the ECAP sample.\\

Upon removing the ECAP sample point, significant changes in correlations are evident, particularly regarding the dislocation density, $\rho$. This variation can be attributed to dislocation annihilation occurring during the initial minute of high-temperature annealing, as detailed in \cref{subsubsec:dislocation_density} and \cref{fig: XRD_data}\textcolor{blue}{(b)}. The sudden initial drop in dislocation density results in the identification of some unusual and unphysical correlations, such as between dislocation density and ($11\bar{2}0/0002$) texture ratio. Additionally, the dislocation density exhibits small positive correlations with precipitate volume fractions. This may arise from the simultaneous occurrence of dislocation annihilation and precipitate dissolution upon annealing, leading to these unexpected correlations.\\

Our analysis in \cref{fig:Correlations}\textcolor{blue}{(b)} reveals a strong negative correlation between grain size ($d$) and precipitate volume fractions, consistent with Zener theory (\cref{Zenereqn}). Furthermore, grain size exhibits a robust negative correlation with texture for the ($11\bar{2}0/0002$) texture ratio, which could explain the observed trend of the texture becoming more basal over time, as depicted in the pole figures in \cref{fig: Microstructural Evolution}\textcolor{blue}{(e)}. Moreover, we observe high correlations among the three precipitate volume fractions, which is unsurprising given their relationship through the constraint that their sum equals 1. This correlation likely arises from the simultaneous dissolution of all these precipitates over time. 

The correlations presented here offer valuable insights into the interplay between various microstructural features in our alloy system, shedding light on the complex relationships governing its behavior. Taken together, the data underscores the importance of accelerated characterization combined with advanced statistical analysis in understanding such correlations.
\subsection{Machine Learning-guided Structure Property Correlations}
\subsubsection{Strengthening Mechanism}
\label{sec: Strength_Mechanism}

In this investigation, we utilize a machine learning approach comprising Pearson correlation coefficient~\cite{freedman2007statistics} and LASSO regression~\cite{tibshirani_regression_1996} analyses to understand the influence of individual microstructural features on hardness. Recognizing the limitations posed by a small dataset, we address this challenge by integrating established physics-based models for strengthening into these feature selection models. This integration involves leveraging physics-based features, a strategy detailed below.\\

In general, the common physical factors contributing to the strengthening of a magnesium alloy are solid solution strengthening ($\sigma_{SS}$), dislocation strengthening ($\sigma_{dis}$), grain size strengthening ($\sigma_{gb}$), and precipitation strengthening ($\sigma_{ppt}$). Based on previously known theory, we can estimate the total strengthening in the alloys to be a sum of individual strengthening mechanisms. \cite{eswarappa_prameela_strengthening_2022}

\begin{eqnarray} \label{eqn:strengthening_mechanism}
    \sigma & = & \sigma_{SS} + \sigma_{Dis} + \sigma_{GB} + \sigma_{ppt}, \nonumber \\
    & = & m\sum_i B_i(X_i)^{2/3} +  M\alpha Gb \sqrt{\rho} + kd^{-1/2} + \Delta \tau. \nonumber \\
\end{eqnarray}

In Equation \ref{eqn:strengthening_mechanism}, the first term represents the contribution from solid solution strengthening as explained by the Labusch model \cite{toda-caraballo_understanding_2014,wang_multi-solute_2022}, the second term represents the contribution from dislocations that can be estimated by \cite{eswarappa_prameela_strengthening_2022}, the third term represents grain boundary strengthening based on the Hall-Petch relationship \cite{cordero_six_2016}, and the last term represents the contribution from precipitates as estimated using the Orowan model. \cite{eswarappa_prameela_strengthening_2022} The individual fits of each of these models are listed in the \ref{SI-StrengthIndividualFits}. \\

For solution-strengthening, $X_i$ denotes the atomic fraction of solute $i$, and $B_i$ represents the potency factor corresponding to solute element $i$. Assuming all the solute atoms ($0.4$ at.\% of Zn and $0.2$ at.\% of Ca) are in solution, the maximum solution-hardening is estimated to be only $10$ MPa / $3.3$ H$_V$ (calculated from \cite{wang_high-throughput_nodate}). We neglect the contribution of Mn due to its minimal content ($<0.07$ at.\%). Since this maximum hardening is barely 1\% of the measured hardness, we consider solid solution-strengthening to be negligible and do not consider it for the analyses. In Equation \ref{eqn:strengthening_mechanism}, the parameter $\alpha = 0.2$ captures dislocation interactions within the basal slip system, $G$ signifies the shear modulus of the Mg matrix (approximately $16.6 \, \text{GPa}$), $\rho$ encapsulates dislocation density, $b$ represents the Burgers vector (roughly $0.32 \, \text{nm}$), $M$ is the Taylor factor ($\approx 4.5$ in Mg alloys \cite{duley_implications_2021}), $d$ signifies the grain size, $k$ is the Hall-Petch slope obtained by fitting our data into the above model, and $\Delta \tau$ is the increment in the critical resolved shear stress (CRSS) due to precipitates. \cite{prameela_rapid_2023} To quantify the strengthening effect of precipitates, the increment in the $\Delta \tau$, resulting from the necessity for dislocations to bypass two distinct precipitates, is estimated as follows:
\begin{equation}
    \Delta \tau = \dfrac{Gb}{2\pi \lambda^* \sqrt{1-\nu}}\ln{\dfrac{d_{p_i}^*}{r_o}}.
\end{equation}
Here, $\lambda^*$ signifies the effective planar inter-particle spacing on the slip plane, $\nu$ is the Poisson's ratio of the Mg matrix ($\approx 0.3$), $d_{p_i}^*$ denotes the mean planar diameter of the particles on the slip plane calculated from XRD using the Scherrer equation and assuming that peak broadening is due to particle size effects. $r_o$ is the core radius of the dislocations, approximated to be the magnitude of $b$ ($r_0 = 0.32 \, nm$). \cite{duley_implications_2020} Given the Scherrer equation is only valid for particle sizes less than 200 nm \cite{holzwarth_scherrer_2011}, we limit our consideration to particles below this threshold, which are the ones most responsible for strengthening. Following the methodology outlined by J.F. Nie \cite{nie_effects_2003}, the mean inter-precipitate spacing can be approximated based on the volume fraction of the precipitates, $f$ and mean diameter of the precipitates, $d_{p_{i}}^*$ as:
\begin{equation}
    \lambda = (\dfrac{0.779}{\sqrt{f}} - 0.785) \times d_{p_{i}}^*.
\end{equation}
The precipitate size and spacing obtained through this route are listed in the \cref{tab: ppt_size_spacing}. We carried out hardness (H$_\text{V}$) to yield strength ($\sigma_y$) conversions using the relationship where ($\sigma_y \approx 3.3 \times \text{H}_\text{V}$). \cite{zhang_general_2011} Given the absence of existing models describing the influence of texture on strength, we opted to utilize the texture intensity ratios derived directly from XRD. \\

\begin{figure*}[h!]
    \centering 
    \subfigure[]{\includegraphics[height=5.5cm,angle=270,origin=c]{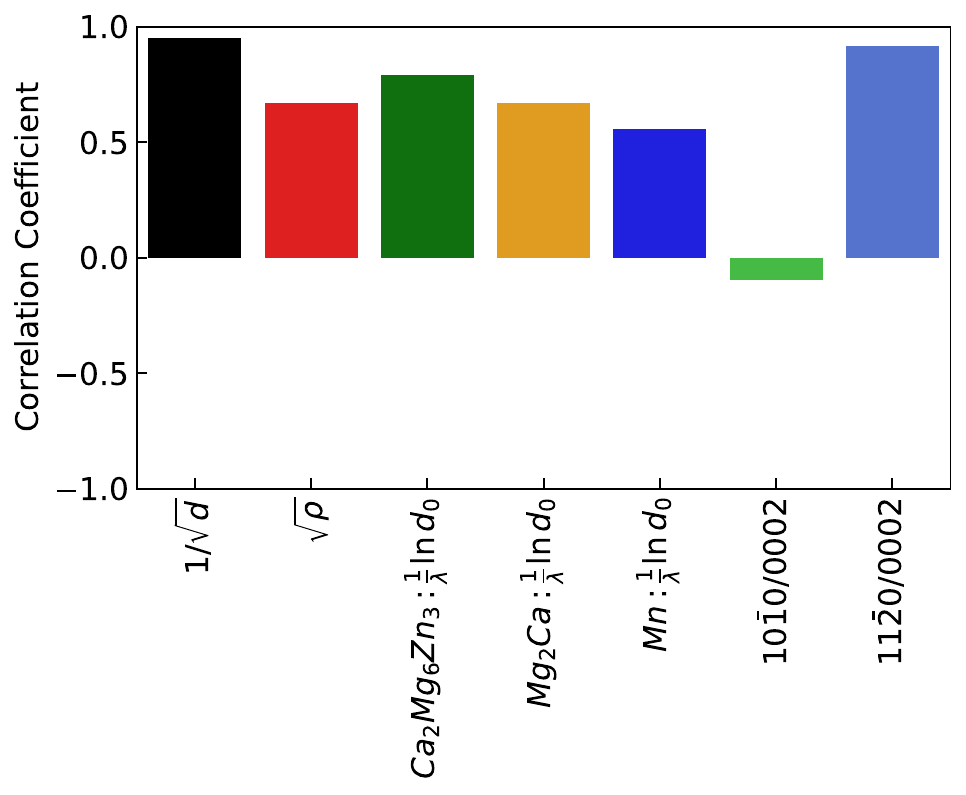} \label{subfig:hardness_pcc}}
   \subfigure[]{\includegraphics[height=5.3cm,angle=270,origin=c]{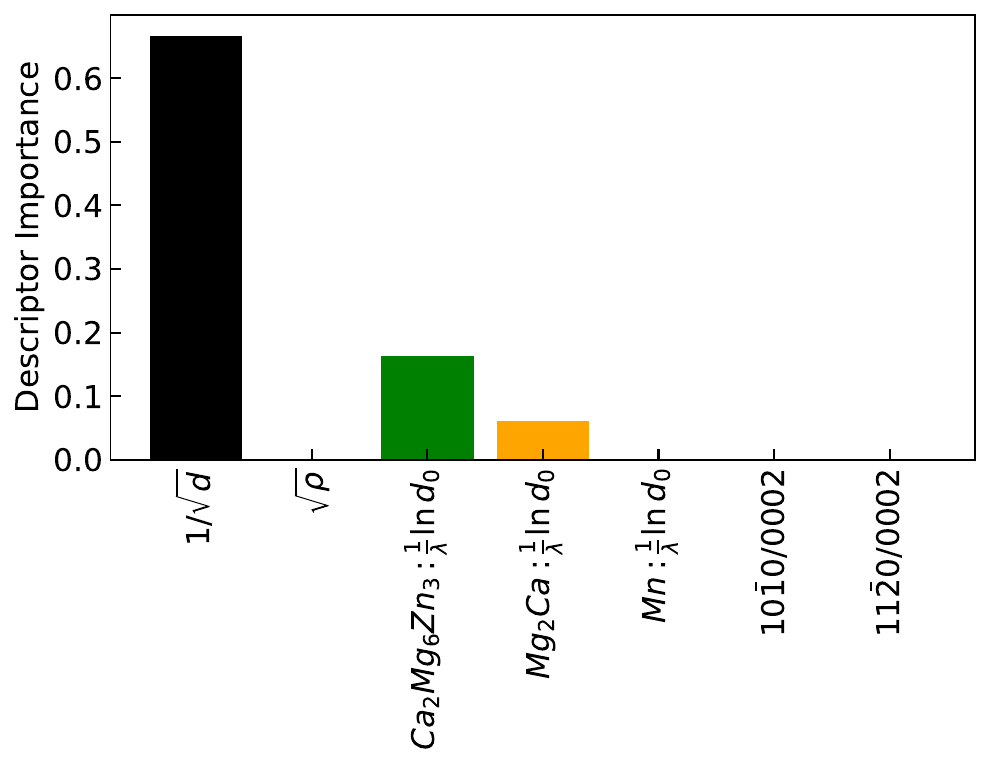} \label{subfig:hardness_lasso}}
    \caption{Identification of dominant microstructural features for strengthening using (a) Pearson R correlation coefficients~\cite{freedman2007statistics} and (b) LASSO~\cite{tibshirani_regression_1996}.}
    \label{fig:hardness_ML_analysis}
\end{figure*}

Using the above relations, we first conducted a correlation analysis using the Pearson R correlation coefficients~\cite{freedman2007statistics} between the microstructural features and the hardness value as seen in \cref{subfig:hardness_pcc}. We observed that the inverse square root of the mean grain size ($1/\sqrt{d}$) is strongly positively correlated with hardness, implying that the hardness value increases with a decrease in the grain size, consistent with Hall-Petch strengthening. \cite{cordero_six_2016}
Likewise, the descriptors for precipitates in both ternary and binary phases exhibit a positive correlation with hardness values, suggesting that the presence of precipitates enhances alloy hardness, as anticipated in precipitation strengthening. \cite{eswarappa_prameela_strengthening_2022,nie_precipitation_2012} This correlation may also be due to precipitates hindering grain growth rather than directly resulting from precipitation strengthening.\\

However, the unusual, highly positive correlation observed with ($11\overline{2}0/0002$) texture ratio and its impact on hardness could stem from its significant decrease over time. Similarly, the very slight negative correlation seen with ($10\overline{1}0/0002$) texture ratio might result from the relatively constant intensity of ($11\overline{2}0$) plane during annealing, coupled with changes primarily in ($0002$) plane. Although both textures exhibit correlations, these are indirect and do not directly affect hardness, as indicated by LASSO descriptors below and previous literature. Instead, it's the precipitates influencing grain growth that effectively impact texture as elaborated in \cref{subsec: texture}.\\

As correlations do not imply causation, we conducted LASSO regression to further understand the importance and dominance of microstructural features. LASSO, a method known for its ability to simplify models by eliminating less impactful variables, provided a refined understanding of the most significant features affecting hardness. Our LASSO model estimated the contribution and dominance of each feature using physics-informed and microstructure-based models, as seen below:
\begin{eqnarray}
    H_V & = & \lambda[A(1/\sqrt{d}) + B\sqrt{\rho} + \sum_i C_i(\frac{1}{\lambda^*} ln[d_0]) + ... \nonumber \\
     & ... & + D\frac{(10\overline{1}0)}{(0002)} + E\frac{(11\overline{2}0)}{(0002)}],
\end{eqnarray}
where A, B, $C_i$, D, and E are fitting constants in LASSO that give the feature importance value. 
The LASSO regression fit has an accuracy of 93.14\%, and the fit was obtained by optimizing over the $\lambda$ parameter using ``leave-one-out''-cross-validation (LOOCV).\\

In \cref{subfig:hardness_lasso}, our analysis underscores the predominant influence of grain size on the mechanical properties of the materials studied, aligning with established literature on the dominance of grain size strengthening in Magnesium alloys. \cite{nussbaum1989strengthening,peng2018effect,wang_grain_2023} This phenomenon arises from the hindrance of dislocation and twinning motion by grain boundaries, which enhances the yield strength within the Mg matrix. The Hall-Petch slope ($k$), ranging from 90 to 300 MPa $\mu m^{1/2}$, surpasses that of Al alloys or Steel by a factor of 2-5, highlighting the significance of grain size in strengthening. \cite{wang_grain_2023,cordero_six_2016}\\

While both ternary and binary precipitates contribute to material strengthening, our analysis suggests that the ternary phase plays a relatively more substantial role in accordance with experimental observations. This could result from the higher volume fraction of the ternary phase over the binary phase. Despite their positive correlation with hardness, the impact of precipitates on strengthening, as inferred from LASSO analysis, appears comparatively modest. This observation aligns with the limited effectiveness of precipitates in strengthening via the Orowan mechanism \cite{nie_precipitation_2012}, likely due to their susceptibility to cutting by dislocations. \cite{wang2020dislocation,wang_grain_2023} Additionally, in the context of a dilute alloy system, the effectiveness of precipitates in strengthening is further compromised. The tendency for coarsening and dissolution of precipitates over time leads to diminished strengthening effects, as confirmed by LASSO analysis. The relative influence of precipitates on strength is largely dictated by their volume fraction, with the ternary phase exhibiting the highest importance, followed by binary precipitates, and then the Mn phase with its extremely low volume fraction. However, it is worth noting that precipitates indirectly contribute to strengthening by impeding grain growth during solution treatment, thereby enhancing strength through the Hall-Petch mechanism. This explains the discrepancy between the high correlation observed through Pearson R Correlation and the relatively lower importance assigned by LASSO analysis.\\
 
In contrast, dislocation strengthening does not feature prominently in our analysis despite its high correlation with hardness. This discrepancy arises from the significant drop in dislocation density during the initial phase of high-temperature solution treatment, followed by stabilization at relatively constant levels. However, lower temperature annealing regimes may reveal a more pronounced impact of dislocations on hardness, as indicated by PCC analysis. \cite{eswarappa_prameela_strengthening_2022} \\

When considering other microstructural features, the contribution of texture to strengthening appears negligible, likely due to its minimal influence on hardness. This observation is consistent with prior research highlighting the limited effect of texture on material hardness. \cite{guo_mechanism_2017,vlassak1994measuring}

\subsubsection{Corrosion Mechanism}

Corrosion rates in Mg alloys, much like strengthening, are influenced by various microstructural factors, including grain size, secondary phases, texture, and dislocation. However, there remains a gap in theoretical models linking these microstructural features to corrosion rates. Only a few studies have attempted to establish such correlations~\cite{ralston_revealing_2010,bahmani_corrosion_2019}.\\

In line with methodologies used for understanding strengthening mechanisms, we employ Pearson correlation analysis~\cite{freedman2007statistics} and LASSO regression techniques~\cite{tibshirani_regression_1996} to assess the influence of individual microstructural features on corrosion rate. To mitigate the limitations posed by a relatively small dataset, we integrate established physics-based models for strengthening into the feature selection process of both Pearson correlation coefficient (PCC) and LASSO regression models, as elaborated below.\\

Grain size ($d$) is one of the most critical microstructural features influencing corrosion rate. Given that grain boundaries harbor more lattice defects and dislocations compared to the interiors of grains, grain boundaries are expected to corrode faster when exposed to corrosive environments. As a result, grain boundaries are considered to accelerate the corrosion rate. However, the relationship between grain size and the corrosion rate is complex, with conflicting reports suggesting that larger or smaller grain sizes may either decrease or elevate the corrosion rates. \cite{chen2017correlation,song_control_2007,song2011corrosion,bahmani_corrosion_2022,bahmani_formulation_2020}\\

Here we utilize the Hall-Petch type model that Ralston and Birbilis proposed to explain the relationship between corrosion rate and grain size~\cite{ralston_revealing_2010}:
\begin{equation}
    CR=A+\dfrac{B}{\sqrt{d}}.
\label{Ralston Model}
\end{equation}
where the constant $A$ depends on the specific environmental conditions, and $B$ represents a material constant that varies with composition. The fit of the Ralston-Birbilis model is listed in Supplementary \cref{fig:RB Model}. A modified version of the Ralston-Birbilis model was proposed by Bahmani \textit{et al.} \cite{bahmani_formulation_2020}, where the grain size effect is limited to the matrix phase by the matrix fraction, ${f_{Mg}}$.
\begin{equation}
    CR = A + B \times \frac{f_{Mg}}{\sqrt{d}}.
    \label{Bahmani Model}
\end{equation}
Lastly, Bahmani \textit{et al. }\cite{bahmani_formulation_2020} combined the effects of the grain size and precipitate effects on corrosion into the following model:
\begin{equation} \label{eqn:corrosion}
    CR = A + B\times \dfrac{f_{Mg}}{\sqrt{D}} + C\times \sum_i f_i|\Delta E_i|.
\end{equation}
where $f_i$ represents the volume fraction of the intermetallic phase obtained from XRD, and $\Delta E_i$ signifies the volta-potential difference between the intermetallic and Mg matrix. The fit of the Bahmani precipitate model is listed in Supplementary \cref{fig:BP Model}.\\

As per the galvanic series, Mg ranks among the most anodic structural metals, whereas most secondary phases, with the exception of Mg$_2$Ca, exhibit cathodic characteristics, rendering Mg more prone to corrosion. \cite{sudholz_electrochemical_2010} The corrosion rate attributed to micro-galvanic cells due to the presence of intermetallic phases can be effectively modeled by considering the voltaic potential difference (obtained from \cite{bahmani_corrosion_2019}), the kinetics of the reaction, and the volume fraction of the phase, which represents the available sites for corrosion. The third term in \cref{eqn:corrosion} establishes the correlation between corrosion and the presence of precipitates. Remarkably, fitting the precipitate model to the corrosion rate data yields an $R^2$ value of 0.87 as shown in Supplementary \cref{fig:RB Model,fig:BP Model}.\\

It has been known that dislocation density and crystallographic texture can also impact corrosion rates significantly. \cite{bahmani_formulation_2020, gerashi_effect_2022} According to the literature \cite{jiang_stress_2023,bahmani_formulation_2020}, excess dislocations often accelerate the nucleation and expansion of pitting and the generation of corrosion products. Texture also significantly affects the corrosion behavior of Mg alloys as the work function is different on different crystallographic planes. \cite{bahmani_corrosion_2022} However, no physical or numerical models exist to describe their contributions, thus facilitating the need to correlate them to corrosion rates directly.\\

\begin{figure*}[htbp]
    \centering
    \subfigure[]{\includegraphics[height=5.5cm,angle=270,origin=c]{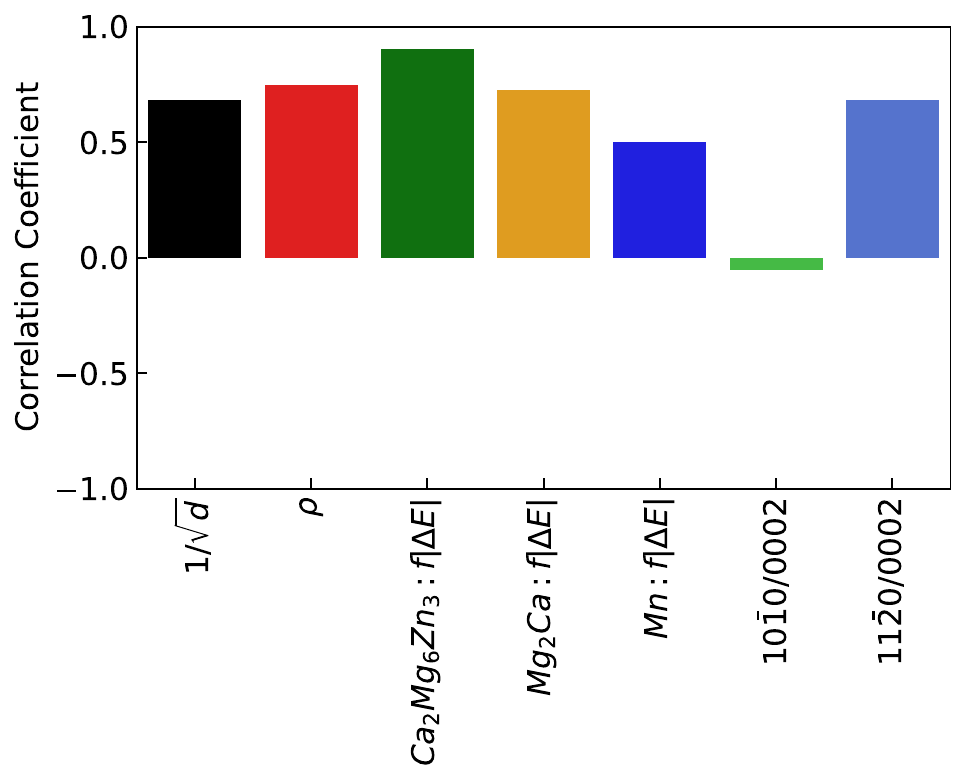} \label{subfig:corrosion_pcc}}
    \subfigure[]{\includegraphics[height=5.4cm,angle=270,origin=c]{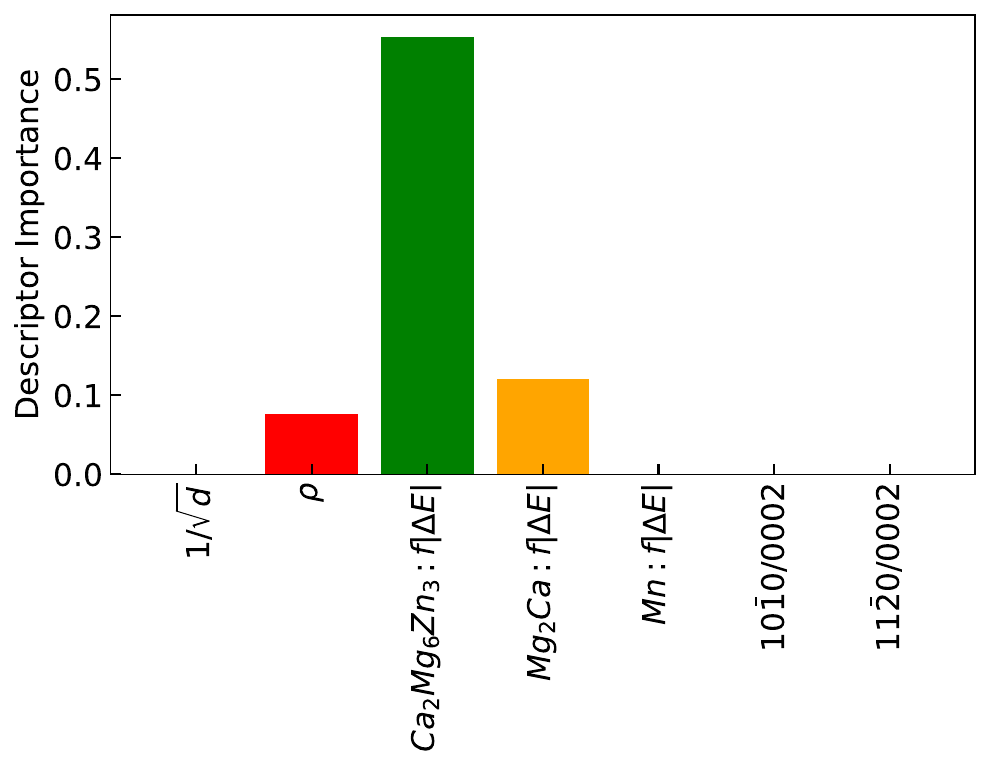} \label{subfig:corrosion_lasso}}
    \caption{Identification of dominant microstructural features influencing corrosion using (a) Pearson R correlation coefficients~\cite{freedman2007statistics} and (b) LASSO~\cite{tibshirani_regression_1996}.}
    \label{fig:corrosion_ML_analysis}
\end{figure*}

To enable this analysis, we first conducted correlation analyses, as shown in \cref{subfig:corrosion_pcc}, to understand the correlation between microstructural features and corrosion rate, combining the physical relationships outlined above. In this data set, we observed positive correlations for dislocation density and precipitate fraction (with ternary being the most correlated), consistent with the literature that dislocation density and precipitates increase the corrosion rates. Additionally, $(11\overline{2}0/0002)$ texture ratio showed a positive correlation, in line with literature \cite{gerashi_effect_2022} suggesting that prismatic planes corrode faster, while $(10\overline{1}0/0002)$ texture ratio exhibited a small negative correlation. However, this slight negative correlation might be attributed to the relatively constant intensity of $(10\overline{1}0)$ plane during annealing, coupled with changes primarily occurring in the $(0002)$ plane. The inverse square root of grain size also showed a positive correlation, suggesting that grain size refinement increases corrosion. However, it's essential to note that a high correlation does not necessarily equate to influencing corrosion behavior.\\ 

To truly distinguish between these microstructure features and gain insights into the dominant microstructure, we conducted LASSO fitting, as outlined below:
\begin{eqnarray}
    CR & = & \lambda[A(1/\sqrt{d}) + B\rho + C \times \sum_i f_i|\Delta E| + ... \nonumber \\
     & ... & + D\frac{(10\overline{1}0)}{(0002)} + E\frac{(11\overline{2}0)}{(0002)}],
\end{eqnarray}
where A, B, C, D, and E are fitting constants in LASSO that give the feature importance value. The LASSO regression fit has an accuracy of 81.16\%, and the fit was obtained by optimizing over the $\lambda$ parameter using LOOCV; see \cref{subsec:methods_ML}.\\

The literature presents an interesting and somewhat contradictory perspective on the relationship between grain size and corrosion rate. Bahmani \textit{et al.} \cite{bahmani_corrosion_2019,bahmani_corrosion_2022} suggest that an increase in grain size may reduce corrosion rate due to factors such as decreased lattice strain, dislocation density, and surface potential. Conversely, a decrease in grain size may lead to a lower corrosion rate due to the formation of a more uniform and coherent passivation layer, increased barrier effects at grain boundaries against crystallographic pitting, and enhancement of basal plane intensity. Despite a positive correlation observed with $1/\sqrt{d}$ in Pearson R correlation analysis, LASSO regression (\cref{subfig:corrosion_lasso}) indicates that grain size may not have a significant impact on corrosion rate for this Mg alloy and under these processing conditions.\\
 
The LASSO analysis, mirroring PCC analyses, reveals that corrosion rates are predominantly influenced by the ternary \ce{Ca2Mg6Zn3} phase. The potential difference between the intermetallic phase and the matrix drives electrons from the anode to the cathode, forming galvanic cells. In this electrochemical process, the anodic second phase degrades upon contact with the nobler matrix in a galvanic cell, while the matrix degrades when the second phase is nobler than the matrix. Specifically, \ce{Ca2Mg6Zn3} acts as a cathode, and the Mg matrix serves as an anode at their interface, enabling the formation of a micro galvanic cell and thereby increasing the corrosion rate. \cite{zhang2008microstructure,hofstetter_high-strength_2014,hofstetter_processing_2015}\\

\ce{Mg2Ca} exhibits a positive correlation with corrosion and contributes to corrosion despite being more anodic than Mg \cite{zhang2008microstructure}. Kim et al. \cite{kim2008influence} demonstrated that the formation of a galvanic cell between the Mg anode and \ce{Mg2Ca} cathode accelerated the hydrogen evolution rate, thereby increasing corrosion. However, their impact is relatively minor due to their lower volume fraction and potential differences compared to the ternary phase.\\

Despite its positive correlation, the Mn phase appears to play no significant role in corrosion behavior. While the potential difference \cite{bahmani_corrosion_2019} indicates that the Mn phase also acts as a cathode, forming a galvanic couple and leading to increased corrosion rates as seen in PCC analyses, its extremely low content in these alloys renders it of minimal impact, as reflected in the LASSO analysis.\\

Generally, dislocations act as anodic sites relative to the matrix \cite{bahmani_formulation_2020}, leading to a strong positive correlation between dislocation density and corrosion rate. This effect becomes most pronounced in highly deformed samples and can even persist following low-temperature thermal treatments in which dislocations can be fairly stable. For the high-temperature anneals performed here, the LASSO analysis indicates that dislocation density has little influence, likely due to the rapid annihilation of dislocations within the first minute of annealing. Further research incorporating a range of dislocation densities is necessary to comprehensively grasp their impact on the corrosion behavior of this Mg alloy. \\ 

Similar to dislocations, literature \cite{gerashi_effect_2022} suggests that texture exerts some influence on corrosion rate. While there is a positive correlation, the LASSO analysis indicates the impact is negligible for high-temperature thermal processing of this Mg alloy. The lack of influence can be partly attributed to the absence of strong texture in this ECAP sample, even following high-temperature anneals. We argue that a relatively random distribution of textures effectively nullifies its impact. 
\section{Conclusions}
In this study, we used rapid characterization techniques and machine learning analyses to investigate the complex interplay between thermal processing, microstructure, hardness, and corrosion rates in an Mg alloy. Utilizing XRD and optical microscopy, we swiftly characterized microstructural features with minimal sample preparation, while hardness measurements and 1-day immersion tests provided rapid insights into mechanical and corrosion behavior, respectively.\\

While the anticipated trend of decreasing hardness and corrosion rates with prolonged high-temperature annealing was observed, intriguing deviations were noted at short annealing times. Particularly, a 1-minute anneal at 450$\degree$C yielded a favorable combination of high hardness and relatively low corrosion rate. This brief annealing significantly reduced dislocation density and dissolved ternary precipitates. Extending the annealing time to two minutes resulted in a considerable increase in grain size accompanied by texture variations. These distinct variations underscore the importance of employing accelerated characterization techniques to capture variations across various annealing durations.\\

Despite the challenges posed by strong correlations among microstructural features and limited data, we successfully established correlations between microstructural features, hardness, and corrosion rates using machine learning-based feature selection routes, such as LASSO regression, as well as Pearson R Correlations in conjunction with physics-based relationships.\\

Our analysis highlights the significant roles of grain size refinement in strengthening and the control of corrosion rates by ternary phase fraction, as revealed by LASSO. Importantly, our findings emphasize that achieving a fine grain size and reducing the presence of ternary phases and dislocations can yield an optimal combination of strength and corrosion resistance. These insights not only enhance our understanding of material behavior but also offer valuable guidance for processing alloys tailored to specific application requirements.
\section{Declaration of competing interest}
The authors declare that they have no known competing financial interests or personal affiliations that might have influenced the findings presented in this paper.
\section{Acknowledgements}
The authors extend their gratitude to Prof. Todd Hufnagel for insightful discussions regarding X-ray diffraction analyses, Dr. Karthikeyan Hariharan for access to the CALPHAD phase diagram, and Mr. Tunde Ayodeji for assisting with graphical illustrations. Additional thanks to Diana Bershadky, Dr. Suhas Eswarappa Prameela, Dr. Jenna Krynicki, Mr. Rich Middlestadt, Dr. John Fite, and Prof. Roger Guillory II for their invaluable insights and support.  
S.R. and T.P.W were supported by the National Science Foundation under grant DMR \#2320355. 
M.S.P. and P.C. were supported by the 
Department of Energy, Office of Science, Basic Energy Sciences, under Award \#DE-SC0022305
(formulation engineering of energy materials via multiscale learning spirals).
Computing resources were provided by the ARCH high-performance computing (HPC) facility, which is supported by National Science Foundation (NSF) grant number OAC 1920103.

\printbibliography

\clearpage
\appendix
\onecolumn

\titleformat{\section}[hang]{\normalfont\large\bfseries}{\thesection}{0.5em}{} 
\titlespacing*{\section}{0pt}{0.5em}{0.05em} 

\titleformat{\subsection}[hang]{\normalfont\bfseries}{\thesubsection}{0.1em}{} 
\titlespacing*{\subsection}{0pt}{0.01ex plus 0.01ex minus 0.01ex}{0.01em plus 0.01em minus 0.01em} 

\titleformat{\section}[hang]{\normalfont\large\bfseries}{\thesection}{1em}{} 
\titlespacing*{\section}{0pt}{0.1em}{0.1em}  

\titleformat{\subsection}[hang]{\normalfont\bfseries}{\thesubsection}{1em}{} 
\titlespacing*{\subsection}{0pt}{0.1em}{0.0em} 

\section{Supplementary Information}
\renewcommand{\thefigure}{SI~\arabic{figure}}
\setcounter{figure}{0}
\renewcommand{\thetable}{SI~\arabic{table}}
\setcounter{table}{0}

\titlespacing*{\subsection}{0pt}{0.0em}{0.0em}

\subsection{Grain Size Measurements}
\begin{table}[H]
        \centering
        \begin{adjustbox}{width=1\textwidth}
        \begin{tabular}{|c|c|c|c|c|c|}
        \hline
\textbf{Condition} & \textbf{Time (hr)} & \textbf{Mean ($\mathbf{\mu}$m)} & \textbf{St.Dev} & \textbf{Minimum ($\mathbf{\mu}$m)} & \textbf{Maximum $\mathbf{\mu}$m)}\\
\hline
ECAP & 0 & 0.83 & 0.35 & 0.33 & 2.11\\
1 min & 0.02 & 0.84 & 0.40 & 0.24 & 2.29\\
2 min & 0.07 & 11.52 & 4.73 & 2.32 & 20.78\\
5 min & 0.08 & 19.39 & 9.97 & 3.14 & 51.34\\
15 min & 0.25 & 18.17 & 12.76 & 3.29 & 66.01\\
30 min & 0.50 & 20.82 & 12.87 & 2.10 & 62.42\\
1 hr & 1.00 & 21.27 & 12.37 & 4.89 & 62.20\\
4 hr & 4.00 & 25.66 & 12.89 & 4.46 & 58.86\\
16 hr & 16.00 & 27.61 & 17.88 & 6.34 & 99.99\\
64 hr & 64.00 & 40.39 & 29.04 & 5.41 & 158.34\\
128 hr & 128.00 & 49.40 & 23.42 & 11.03 & 137.49\\
     \hline
        \end{tabular}
        \end{adjustbox}
        \caption{Grain Size Statistics}
        \label{tab: Grain Growth Variations}
    \end{table}
\subsection{Normalized Grain Size Distributions}
\begin{figure}[H]
\centering
\includegraphics[width=1\linewidth]{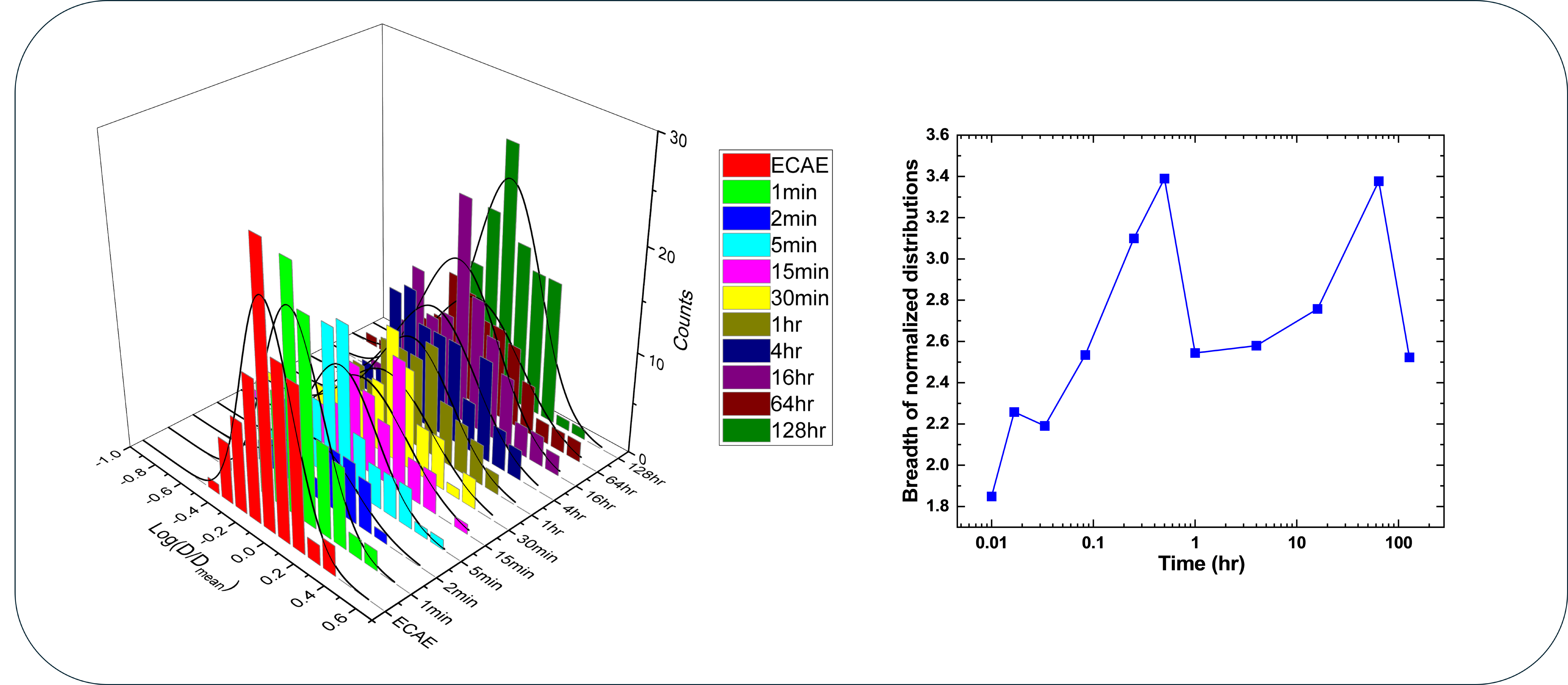}
    \caption{Normalized Grain Size Distributions}
    \label{fig: N-GG}
    \end{figure}
\subsection{Goodness of Log-normal fit}
We can see log-normal distributions throughout, except in the 2-minute scenario where $\chi^2< 0.05$.

\begin{table}[H]
        \centering
        \begin{tabular}{|c|c|c|c|}
        \hline
\textbf{Condition} & \textbf{Time} &{\textbf{Breadth}} & $\mathbf{\chi^2}$ \\
\hline
ECAP   & 0.00   & 1.85 & 0.12 \\
1 min  & 0.02   & 2.26 & 0.89 \\
2 min  & 0.03   & 2.19 & 0.02 \\
5 min  & 0.08   & 2.53 & 0.21 \\
15 min & 0.25   & 3.10 & 0.48 \\
30 min & 0.50   & 3.39 & 0.19 \\
1 hr   & 1.00   & 2.54 & 0.66 \\
4 hr   & 4.00   & 2.58 & 0.26 \\
16 hr  & 16.00  & 2.76 & 0.88 \\
64 hr  & 64.00  & 3.38 & 0.83 \\
128 hr & 128.00 & 2.52 & 0.53 \\
\hline
        \end{tabular}
        \caption{Log-normal grain size distribution}
        \label{tab:Fit-Log-normal}
    \end{table}
\subsection{Geometrically Necessary Dislocation Maps}
The GND maps obtained from EBSD revealing 
\begin{figure}[H]
\centering
\includegraphics[width=13cm]{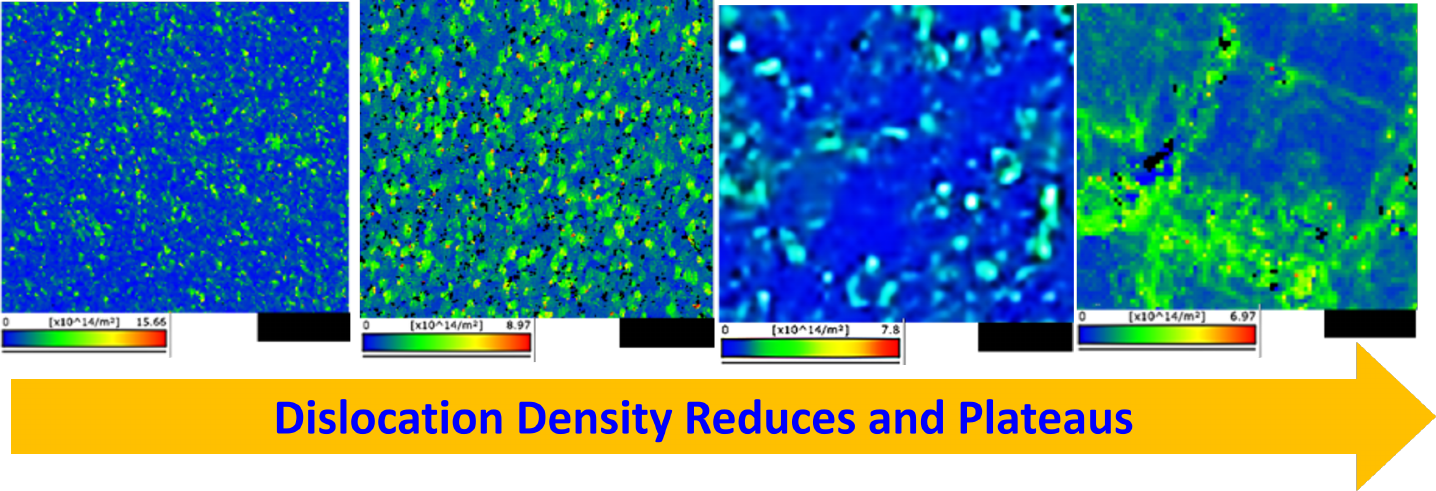}
    \caption{GND Distribution Maps}
    \label{fig: GND_Maps}
    \end{figure}
\subsection{Precipitate Size and Spacing}
\begin{table}[H]
\centering
\resizebox{\textwidth}{!}{%
\begin{tabular}{|c|c|c|c|c|c|c|}
\hline
\multicolumn{1}{|c|}{\multirow{2}{*}{\textbf{Condition}}} & \multicolumn{3}{|c|}{\textbf{Mean Precipitate Diameter (nm)}} & \multicolumn{3}{|c|}{\textbf{Mean Particle Spacing   (nm)}} \\
\cline{2-7}
\multicolumn{1}{|c|}{} & \multicolumn{1}{|c|}{\textbf{\ce{Ca2Mg6Zn3}}} & \multicolumn{1}{|c|}{\textbf{\ce{Mg2Ca}}} & \multicolumn{1}{|c|}{\textbf{\ce{Mn}}} & \textbf{\ce{Ca2Mg6Zn3}} & \textbf{\ce{Mg2Ca}} & \textbf{\ce{Mn}} \\
\hline
\textbf{ECAP} & 149.30 & 117.58 & 131.00 & 277.77 & 374.66 & 473.76 \\
\textbf{1 min} & 147.81 & 129.50 & 148.07 & 345.53 & 432.92 & 663.54 \\
\textbf{2 min} & 158.73 & 116.73 & 214.17 & 385.68 & 395.52 & 965.69 \\
\textbf{5 min} & 146.14 & 146.83 & 161.08 & 357.84 & 486.68 & 746.53 \\
\textbf{15 min} & 151.20 & 109.10 & 104.77 & 373.95 & 368.60 & 462.26 \\
\textbf{30 min} & 150.91 & 194.96 & 168.10 & 373.59 & 651.65 & 759.71 \\
\textbf{1 hr} & 137.44 & 132.64 & 153.98 & 293.44 & 363.97 & 518.52 \\
\textbf{4 hr} & 157.93 & 136.20 & 114.55 & 343.34 & 395.16 & 382.39 \\
\textbf{16 hr} & 131.87 & 180.60 & 171.67 & 298.72 & 531.36 & 606.25 \\
\textbf{64 hr} & 126.42 & 129.02 & 168.75 & 335.67 & 386.09 & 677.80 \\
\textbf{128 hr} & 175.17 & 172.12 & 130.17 & 501.92 & 647.61 & 523.19 \\
\hline
\end{tabular}%
}
\caption{Precipitate Size and Spacing estimated through XRD}
\label{tab: ppt_size_spacing}
\end{table}
\subsection{Strengthening Mechanism Models}
\label{SI-StrengthIndividualFits}

\subsubsection{Hall Petch Model}

The Hall Petch model ($H_V=kd^{-0.5}$) exhibited a superior fit of 89.4\%, denoting a dominant influence of grain boundaries on strengthening.
\begin{figure}[H]
\centering
\includegraphics[width=0.6\linewidth]{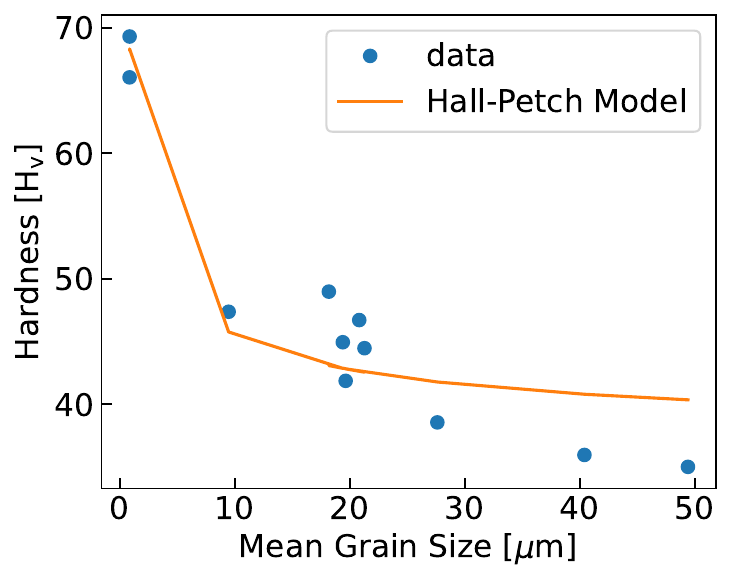}
    \caption{Hall Petch Strengthening}
    \label{fig: Hall_Petch_Plots}
    \end{figure}

\subsubsection{Taylor Dislocation Model}

The poor fit ($R^2 = 0.45$) of the Taylor dislocation model ($\sigma_{dis}=M\alpha Gb\sqrt{\rho}$) represents the suppressed roles of dislocation density in strengthening this material.
\begin{figure}[H]
\centering
\includegraphics[width=0.6\textwidth]{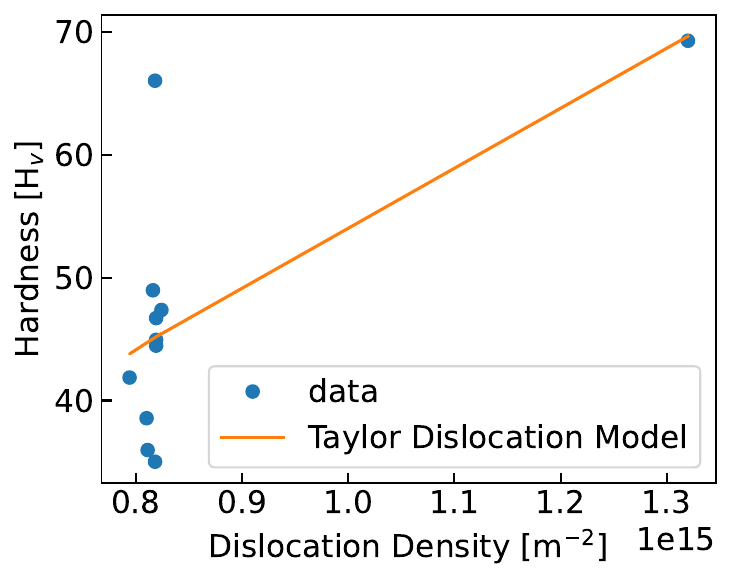}
    \caption{Dislocation Strengthening}
    \label{fig: Taylor_Model_Plots}
    \end{figure}

\subsubsection{Orowan Strengthening Model}
The poor fits of the Orowan model upon strengthening from the ternary ($R^2 = 0.625$), binary ($R^2 = 0.4715$), and Mn ($R^2 = 0.34$) denote the limited roles of precipitates in strengthening this material.
\begin{figure}[H]
 \centering
    \subfigure[]{\includegraphics[width=0.45\linewidth,origin=c]{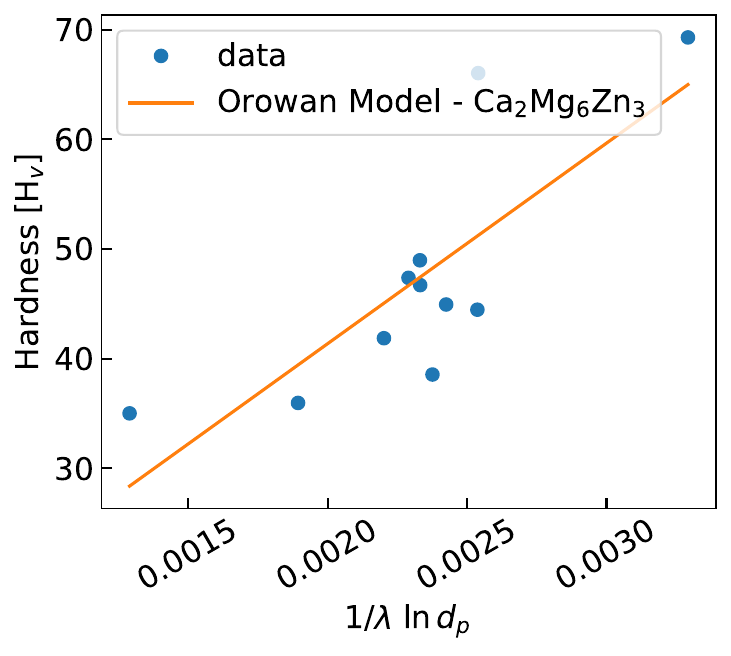} \label{subfig:hardness_ternary}}
    \subfigure[]{\includegraphics[width=0.45\linewidth,origin=c]{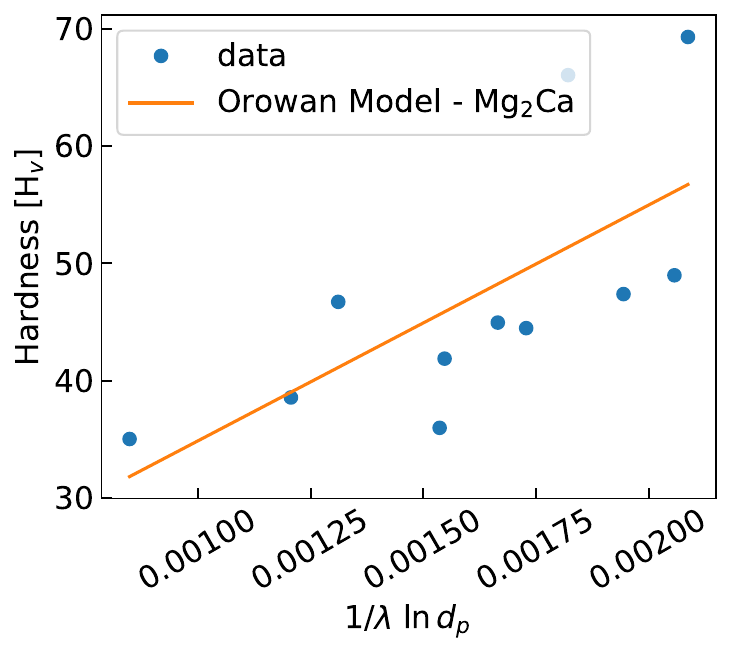} \label{subfig:hardness_binary}}
    \subfigure[]{\includegraphics[width=0.45\linewidth,origin=c]{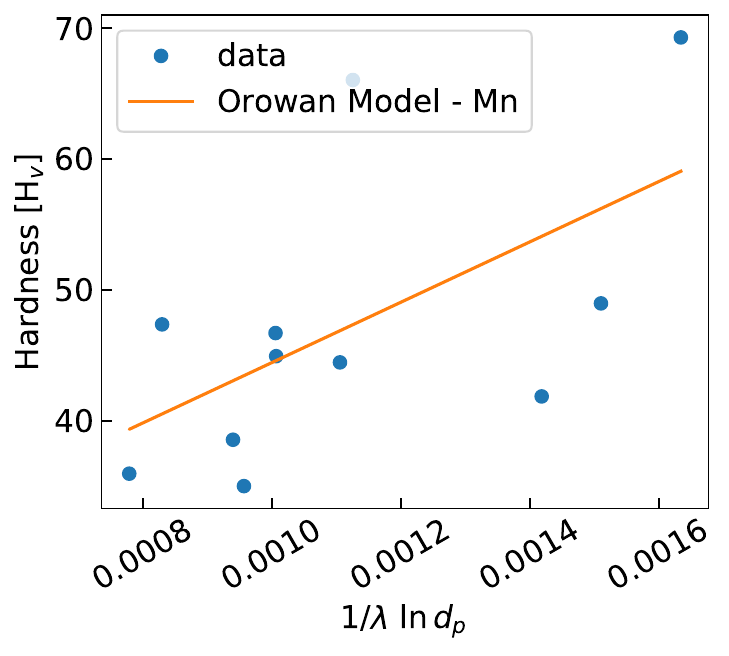} \label{subfig:hardness_Mn}}
    \caption{Precipitate Strengthening by Orowan Mechanism}
    \label{fig:Precipitate_analysis}
    \end{figure}

\subsection{Corrosion Mechanism Models}
\label{SI-CorrosionIndividualFits}
\subsubsection{Ralston-Birbilis Model}
We fitted our grain sizes to corrosion using the Ralston-Birbilis model elaborated in \cref{Ralston Model}. We found a poor fit of 46\%, denoting the suppressed role of grain size in corrosion behavior.

\begin{figure}[H]
    \centering
    \includegraphics[width=0.45\linewidth]{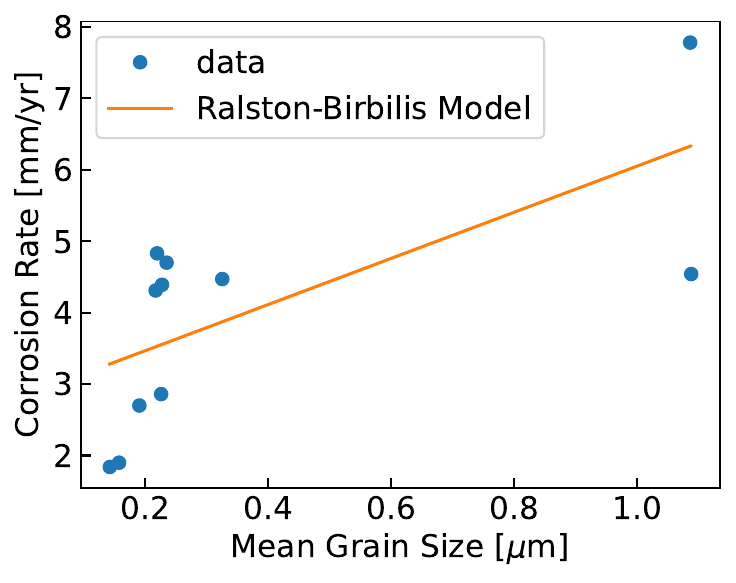}
    \caption{Ralston-Bibilis Grain Size Model - as a function of solution heat treatment}
    \label{fig:RB Model}
\end{figure}

\subsubsection{Bahmani Model}

We fitted our precipitate contributions to corrosion using the Bahmani precipitate model, as shown below:
\begin{equation}
    CR= A + C\times 
    [f_{\text{Ca}_2\text{Mg}_6\text{Zn}_3}|\Delta E_{\text{Ca}_2\text{Mg}_6\text{Zn}_3}|+
    f_{\text{Mg}_2\text{Ca}}|\Delta E_{\text{Mg}_2\text{Ca}}|+
    f_{\text{Mn}}|\Delta E_{\text{Mn}}|
    ]
\end{equation}

We observed a superior fit of 88\%, denoting the dominance of the precipitates on the corrosion behavior.

\begin{figure}[H]
    \centering
    \includegraphics[width=0.45\linewidth]{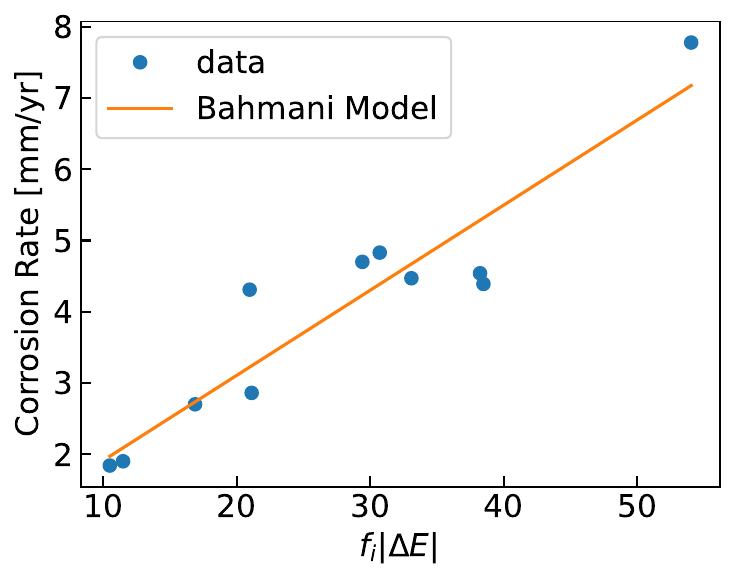}
    \caption{Bahmani Precipitate Model - as a function of solution heat treatment}
    \label{fig:BP Model}
\end{figure}

\end{document}